# Length-scales and fractal scaling in dairy gels: comparisons of the insights from ultra-small angle neutron scattering and rheology

Koduvayur A. Ramya,[1] Liliana de Campo[2], Markus Strobl[3], François Boué[4], and Christopher J. Garvey[1,5]*

1. Technische Universität München, Forschungs-Neutronenquelle Heinz Maier-Leibnitz and Physik Department E13, Lichtenbergstraße 1, 85747 Garching, Germany
2. Australian Centre for Neutron Scattering, Australia Nuclear Science and Technology Organisation, Lucas Heights 2234, Australia
3. Laboratory for Neutron Scattering and Imaging, Paul Scherrer Institut, Villigen, Switzerland
4. Laboratoire Léon Brillouin, CNRS-UMR12-CEA-Université Paris-Saclay, 91190 Gif-sur-Yvette Cedex, France
5. Department of Physics, School of Science, RMIT University, Melbourne, Australia

*Corresponding Author: christopher.garvey2@rmit.edu.au

**Abstract**

A key structural parameter that dictates the textural properties of a particulate gel is the fractal dimension ($D_F$) of the network indicating self-similarity within the spatial range defined by the correlation length ($\xi$). In this study dairy gels are formed from reconstituted skim milk at defined CM volume fractions via two different mechanisms: acidification, and enzymatic (rennet)-induction. Oscillatory rheology is performed to obtain the gelation times, the time-scale for the maturation of mechanical properties of the gel, and the viscoelastic properties of the gel. A direct probe of gel structure is ultra-small-angle neutron scattering (USANS) examining length-scales from approximately 60 nm up to 15 μm. Structural information is extracted from USANS of matured gels by comparing the measured differential scattering cross-section with that calculated from a model constrained by the sample composition. Time-scales for structural equilibration from USANS perspective were shorter than the rheological gelation time for both gels. Rheological scaling theories are utilized to relate the microstructure to their bulk mechanical properties. We show that the fractal scaling at low-$q$ (scattering vectors), due to aggregation of CM, extracted from the USANS at the longest timescales is described by a power-law density-density correlation function, and that rheology relates – through Shih-Wu-Morbidelli models – the local elasticity both within and between the fractal flocs formed by these particles over different size ranges to the bulk moduli of the gels. There are similar values of fractal dimensions from both techniques but different sizes attributed to the complex fractal evolution during dairy gelation, which leaves space to refinements of the models.

## 1. Introduction

The mechanical properties of dairy gels have been interpreted within the general framework of colloidal gels.[1,2] As a stable colloidal dispersion, milk flows but when the colloidal constituents i.e., casein micelles (CM's) undergo an aggregation process, clusters form and percolate into a well-developed network exhibiting a finite elastic modulus, resisting the flow. While the building blocks of the network in different kinds of dairy gels remain the same (the casein micelles, CM), the mechanisms of gelation and corresponding structural changes have a profound effect on the physical properties of the gel.[1,2] The final structure of this gel network has important implications for the gel strength.[3] The network and its structure has been directly observed by small angle scattering (SAS) combined with analysis using simple models[4-6], and inferred from rheological measurements.[1, 2] In both cases, the structural observations are entirely statistical and 3-dimensional since the volume of sample probed is of the order of mL's. The perspectives are non-destructive for SAS, and possibly for rheology at low deformations. SAS has been used to examine the structure of milk, of materials reconstituted from milk powder and of the gels they can form[5-17]. The perspective of small angle neutron scattering, depending on the angle/scattering vector of the normalized intensity, measures the heterogeneity within CM's, or on heterogeneities which make up the gel network.

In general, the initial stages of colloidal gelation are understood as a problem in colloidal stability where the intrinsic stability of the sol is perturbed by a change in the interactions between the constituent particles. The ensuing aggregation process, rather than leading to the precipitation of aggregates, often results in the formation of a network between the aggregates. This occurs when the clusters can quickly diffuse through the surrounding medium to aggregate upon collision.[18] While this structural evolution can be specific to the chemistry of colloidal particles, their interactions, and the percolation kinetics, the final gel may be described as a collection of aggregates called 'fractal objects'.[19, 20] The mass distribution (*M*) within this object will scale with the characteristic cluster size (*r*) via a non-integral power law exponent $D_F$, the fractal dimension, viz., $M(r) = \left(\frac{r}{r_0}\right)^{D_F}$, where $r_0$ is the primary particle size.[19, 20] This fractal object exhibits self-similarity over certain spatial window, *i.e.,* no fractal scaling of the scattering is observed below an inverse maximum size ($1/\xi$), nor beyond an inverse elementary size (1/a).

Colloidal fractals are known to exhibit $D_F$ values between 1 and 3 depending on the aggregation pathway.[19] Mechanisms include fibrillation, diffusion limited cluster aggregation (DLCA, $D_F$ < 2), reaction limited cluster aggregation (RLCA, $D_F$ > 2), shear, and coalescence with the fractal dimension providing an important perspective on the type of aggregates, and the nature of the aggregation process.[20] Although milk is a relatively complex mixture, it is essential to selectively visualize the three-dimensional structure adopted by the colloidal-scale unitary particles, i.e., the casein micelles, responsible for the gel network.[8, 21] The casein micelle (CM) is a heterogeneous particle consisting of water molecules, calcium phosphate, minerals, and various casein proteins. Previous SANS studies focused on length-scales corresponding to the heterogeneity with the CM are consistent with this composition.[8, 9, 22] The CM is a loose aggregated nanostructure defined by an average size and varying degrees of polydispersity.[22, 23] From a colloidal stability perspective, the colloidal sol of CMs is stabilized by interactions between the pendant κ-casein chains, either from electrostatic repulsion when they are charged, and/or by steric repulsion.[24, 25] The addition of a coagulant, acid or enzyme, decreases these repulsions leading to the aggregation of CMs.[1, 2] This is particularly important in the processing of milk gels (e.g., for yoghurt, cheese making) but also in the case of food formulations where textural properties are paramount in consumer appreciation.[26] In this work we have investigated the length scales of the unitary particles, and the fractal nature of the network they form, using ultra small angle neutron scattering (USANS) measured by a neutron based implementation of the Bonse-Hart double crystal diffractometer.[27] This covers a range of length-scales from 10 nm – 10 μm non-destructively over the sample volume probed by the beam-size (~cm$^{-3}$).

Rheology, on the other hand, measures the mechanical properties of the gels. Rheological scaling models consider the strength of a colloidal gel arising from the elasticity of local microstructural constituents[28], neglecting the effect of suspending medium.[3] Incorporating percolation theory, and conformation of the ‘stress-bearing particulate strands’ in the colloidal network enables models to obtain the scaling between moduli and particle volume fraction.[3, 29] These models were first developed around the gelation threshold analogously to the crosslinking of polymer gels.[29] Later, the post-gelation state was described by complementary models for cluster-cluster aggregation of particulate gels, where both the inter-aggregate and intra-aggregate elasticity are equally important.[3, 30] Such a micro-elastic framework has been shown to be useful in rationalizing the link between structure and mechanical properties of dairy gels.[31]

In the present work, we explore the fractal insights from both USANS and rheology experiments to understand the scaling behavior of dairy gels and describe the underlying structural arrangement of CMs resulting from the coagulation process. Both techniques probe bulk sample properties: USANS is well established as a direct structural probe, while the phenomenology of rheological investigations and their subsequent structural rationalization is a standard approach in the food industry.

## 2. Materials and Methods

### 2.1 Materials

The $D_2O$ was used as received from Sigma-Aldrich Chemie (Taufkirchen, Germany). Skim milk powder, specifically the product 21447, was procured from Groupe Lactalis (Laval, France). The composition of this powder is summarized from the corresponding product data sheet in Table 1. Solutions for USANS measurements were prepared by dispersing appropriate amounts of skim milk powder in $D_2O$. The resulting dispersions were magnetically stirred at 30°C temperature in sealed containers for approximately 1-2 hours to produce homogenous optically opaque solutions. For rheology measurements, the skim milk powder at different concentrations, 3.4 to 30 % (w/w), was rehydrated in deionized water as well as in $D_2O$. Sodium azide ($NaN_3$), also procured from Sigma-Aldrich Chemie (Taufkirchen, Germany), was added to the solvent at 0.1 wt. % to prevent microbial contamination of milk.

**Table 1.** Composition of the skim milk powder .

| **Constituents** | **Mass %** |
|---|---|
| Lactose | 53 |
| Proteins (Casein ~ 80%, Whey ~ 20% w/w) | 34 |
| Fat (Phospholipids) | 1.5 |
| Minerals, i.e. Colloidal calcium phosphate (CCP) | 8.5 |
| Moisture | 3 |

Dairy gels were produced by either chemical means, i.e. acid coagulation, or by the action of enzyme on the reconstituted milk. For enzymatic coagulation, the commercial chymosin enzyme CHY-MAX® (Chr. Hansen, Hoersholm, Denmark) at 200 U/ml was added following the stoichiometry of 600 μl coagulant per 40g of reconstituted milk. For acid coagulation, 2% by weight of glucono-δ-lactone powder (Sigma-Aldrich Chemie, Taufkirchen, Germany) was added to the reconstituted milk. For rheology experiments, the coagulant concentration remained the same, while the concentration of milk powder in the reconstituted milk varied. In both cases the addition of the coagulant to the reconstituted milk defined the time t = 0 with complete mixing defining the major uncertainty in the time axis. We estimate this uncertainty to be of the order of seconds.

### 2.2 USANS Measurements

KOOKABURRA[32] is a Bonse-Hart double crystal diffractometer[27] USANS instrument at the Australian Centre for Neutron Scattering (Australian Nuclear Science and Technology Organisation, Lucas Heights, Australia). The Bonse-Hart approach to the analysis of scattered intensity allows evaluation of slit-smeared scattering cross-sections at ultra-small values of the scattering vector, q, where $q=4\pi/\lambda \sin(\theta/2)$, $\lambda$ is the wavelength of neutrons and $\theta$ is the scattering angle. The KOOKABURRA instrument is able to access extremely small angles with excellent resolution in the horizontal direction and poorer vertical resolution (slit smearing[33]) to achieve an effective range of scattering vectors $4 \times 10^{-5} < q < 0.01$ Å$^{-1}$ with a neutron wavelength of 4.74 Å. After a quintuple bounce path through an initial channel cut Si (111) crystal, data points are measured individually as a function of the rotation angle $\theta$ of a second identical analyzer crystal, leading to a so-called rocking curve, a point by point measurement for each q-value. Two types of scans (from low to high q) have been utilized in this study: a full scan of an equilibrium structure which takes 2-4 hours; and a fast scan which covers the first half part of the rocking curve, which in this case takes ~1 hour. The latter procedure was used to efficiently evaluate the approach of scattering curves to their equilibrium configuration after which the longer scan was made.

If the sample evolves during the measurement, as for kinetic studies, the point-by-point nature of the rocking curves is a serious limitation of the USANS technique and careful consideration must be given to the time-scale of the gelation process, at the length-scales under consideration, compared to the time which is required to acquire reasonable counting statistics above the background signal.[34] Additionally, at the very lowest scattering angle interrogated by this type of instrument, it is important to monitor the position of the zero angle with an initial scan of the rocking curve and automatic determination of the relative $\theta = 0$ position. In general, the counting times for the lower angles (or values of q) for gelling systems with a strong power-law decay are of the order 10's of seconds for lowest angles and 10's of minutes for the highest values.[34] The rocking curves were background subtracted (empty cell scattering) and converted to slit smeared scattering cross-sections in absolute scale, $I_{slit}(q)$ using Python scripts specially designed for Gumtree, an open source scientific workbench,[35] based on data reduction methods of Kline[36] for the neutron Bonse-Hart type diffractometer.

Both enzymatic and chemical gelation were performed *in-situ* in the KOOKABURRA sample cells, which comprise a pathlength of 1 mm between two fused quartz windows of 5 cm diameter (~800-1000 µl of sample). The large beam size (~ 10 $cm^2$) of this instrument provides convenient means of increasing the scattered intensity if sufficient sample is available to cover the incident neutron beam, which is the case for reconstituted milk and the resultant dairy gels. Care must be exercised in the contrast and sample thickness to avoid multiple scattering, which would make the analysis of scattering curves problematic.[37] All gelation processes were performed at 30°C, and the USANS is monitored continuously until no further changes in $I_{slit}(q)$ could be observed and analyzed according to the absence of multiple scattering. In order to avoid unwanted microbial contamination, all sample cells were carefully washed, rinsed with ethanol and then dried before the solutions for chemical and enzymatic gelation were introduced.

We did not de-smear the $I_{slit}(q)$ further to calculate the scattered intensity, $I(q)$, obtained with conventional pinhole SANS instruments. As is standard, we analyzed data by comparing the data with a scattering curve computed on the basis of an optimized and appropriate structural model of the sample system, and then convolved calculated model with the instrumental resolution (hence mathematically "smeared"), $I_{model}(q)$.[38] The instrument resolution function was implemented within the software SASVIEW[39] as a vertical smearing $\Delta q_z = 0.0586$ Å$^{-1}$ (vertical $q$ resolution[32]) to the model calculation. The best fits were achieved by minimizing the difference between the measured $I_{slit}(q)$ and the smeared computed $I_{model}(q)$ using a non-linear least squares routine within SASVIEW.

The expected scattering intensity $I_{model}(q)$ of reconstituted skim milk in $D_2O$, before gelation, was computed using a model consisting a form factor, $P(q)$, of a solution of polydisperse spheres (log-normal distribution) and an appropriate structure factor, $S(q)$, for the concentration dependent correlations between spheres[40]:

$$I_{model}(q) = \phi . (\delta_{particle} - \delta_{solvent})^2 . P(q) . S(q) + B_{inc} \qquad (1)$$

where $\phi$ is the volume fraction of particles (here, CMs, $\phi_{CM}$), $B_{inc}$ is the incoherent scattering background, and δ is the scattering length density (SLD) of a particular phase, i.e. CM or solvent, which is a function of the volume averaged isotopic composition for that phase.[41] Although the reconstituted skim milk is a complex mixture (see Table 1), the contrast term $(\delta_{particle} - \delta_{solvent})$, i.e., the difference between the scattering length densities of the constituents approximated by two phases: $\delta_{particle}$ for particles/ CMs, for which the value of has been computed from the internal composition of CMs and the isotopic composition of the water used to make the skimmed milk from powder in compliance with literature[9, 42, 43]; and $\delta_{solvent}$ which at the length-scales explored here is assumed as due to a single phase comprising of the other solutes in the system. The major contributors to the scattering length density and volume fraction considered for the $\delta_{solvent}$ in the model for scattered intensity (Eqn. 1) are shown in the following expression:

$$\delta_{solvent} = \delta_{whey}\phi_{whey} + \delta_{D_2O}\phi_{D_2O} + \delta_{lactose}\phi_{lactose} + \delta_{H_2O}\phi_{H_2O} \qquad (2)$$

For whey protein component of skim milk, we calculated $\delta_{whey}$ with the known structure of ß-lactoglobulin (UniProt P02754) using the program MULCH.[44] Underpinning this approximation is an assumption that all the whey proteins collectively have an equivalent density and composition. A similar approach was used for the lactose component, $\delta_{lactose}$, by specifying the molecular formula of lactose and MULCH. All relevant quantities are thus listed in Table 2.

**Table 2:** The scattering length density, and bulk density of various components of the reconstituted skim milk.

| Constituents | δ ($10^{-6}$ Å$^2$) | Density (g.cm$^{-3}$) |
|---|---|---|
| casein micelle | 3.08[42, 45] | (4.0[46])[b] |
| whey/ ß-lactoglobulin [a] | 2.96 | (1.34)[c] |
| lactose[a] | 3.52 | (1.38)[c] |
| $H_2O$ | -0.50 | 1.00 |
| $D_2O$ | 6.34 | 1.11 |

a. Calculated using MULCH[44] assuming appropriate exchangeable H's per molecule.

b. Rather than density volume fraction of CM's calculated from Eqn. 3.

c. Partial molar volumes from MULCH used for volume fraction calculation.

For the desired concentration of reconstituted skim milk $c_{milk}$, the concentration of total proteins $c_{protein}$, and the concentration of CMs $c_{CM}$, can be estimated from the known composition of skim milk (see Table 1). The volume fraction of CMs, $\phi_{CM}$, in the solution is then calculated using:

$$\phi_{CM} = c_{CM}\left(\frac{g}{ml}\right) \times \text{voluminosity}\left(\frac{ml}{g}\right) \tag{3}$$

where voluminosity is a measure of the volume of solution occupied by CMs per gram of dry casein protein added to the solution. For the present study, we use a value of 4 mL/g.[47] The values of $c_{protein}$, $c_{CM}$, and $\phi_{CM}$ are thus calculated from $c_{milk}$, and reported in Table 3.

**Table 3:** The concentration of total proteins, $\boldsymbol{c_{protein}}$ , and casein micelles $\boldsymbol{c_{CM}}$ **~ 80%** $\boldsymbol{c_{protein}}$ and volume fraction of casein micelles in reconstituted skim milk, $\phi_{CM} = 4.$ $\boldsymbol{c_{CM}}$ (Eq. 3) at various concentrations explored for present work.

| $c_{milk}$ (% w/w.) | $c_{protein}$ (g/L) | $c_{CM}$ (g/L) | $\phi_{CM}$ |
|---|---|---|---|
| 3.4 | 12 | 9 | 0.037 |
| 5 | 17 | 14 | 0.054 |
| 7.5 | 26 | 20 | 0.082 |
| 10 | 34 | 27 | 0.109 |
| 15 | 51 | 41 | 0.163 |
| 20 | 68 | 54 | 0.218 |
| 25 | 85 | 68 | 0.272 |
| 30 | 102 | 82 | 0.326 |

## 2.3 Rheology

The rheological measurements were performed using MCR 702 rheometer (Anton Paar GmbH, Graz, Austria) with a cone & plate geometry (50 mm diameter, and 1° angle of truncation). The rheology measurements were made over a time period of 100 s to 10000 s (~3 hours), comparable with 200 sec (30 min) to 12400 sec (240 min) for the USANS measurements (see scheme below).

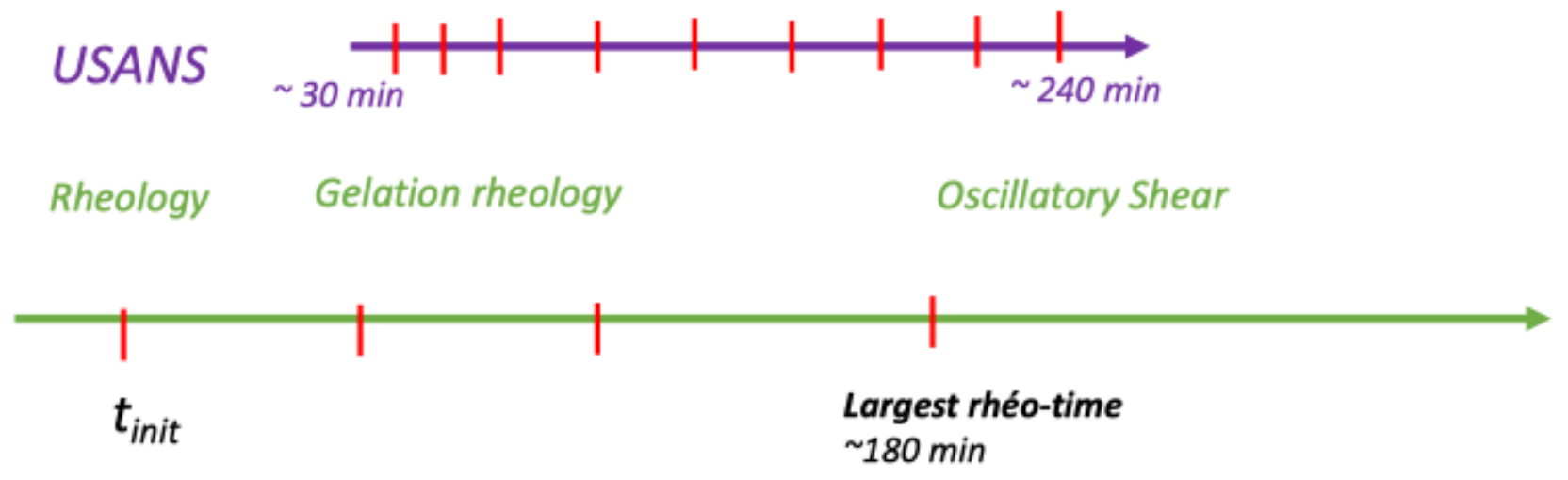

**Figure 1:** Scheme for comparison of time-scales of rheology and USANS investigations.

The cone and plate geometry utilizes relatively low sample volumes within the default measuring gap to obtain the rheological properties accurately during *in-situ* milk coagulation. A temperature of 30 °C was maintained throughout the measurements via Peltier control. A solvent trap enclosing the geometry/ gap, and a low viscosity silicone oil around the rim of the geometry, were utilized to avoid solvent evaporation during measurements. After mixing (~30s), approximately 700 µl of the reconstituted milk-coagulant mixture is transferred onto the bottom Peltier of the rheometer, which is followed by lowering the top geometry, and sample trimming for subsequent testing. Uncoagulated reconstituted skim milk at the same concentrations was also measured independently to confirm the initial Newtonian plateau. The measurements were repeated in $D_2O$ as the solvent, to check for possible deviations in the rheological response compared to the reconstituted skim milk made in deionized water. Two measurement protocols were utilized for each sample:

A. **Gelation Rheology:** Time sweeps in oscillatory shear mode (i.e., strain input, $\gamma = \gamma_0 \sin(\omega t)$) were performed at a constant low strain amplitude ($\gamma_0 = 1\%$) and fixed angular frequencies ($\omega$ = 3.14, 6.28, 31.4 rad/s). The resulting elastic/ storage modulus ($G'$) and viscous/ loss ($G''$) modulus are measured as a function of time to characterize the gel point, and the matured gel. The corresponding loss/ damping factor is given by $\tan\delta(t) = G''(t)/G'(t)$. The mutation time[48] of the system, $t_{mu} = (dlnG'/dt)^{-1}$, defined as the time required for a 1/e-change in its property, here $G'$, at an instantaneous rate of change, $\partial G'/\partial t$, is computed for such an experiment, particularly from the region around the gel point and beyond.

B. **Oscillatory shear sweep:** After gelation rheology to access the gel formation, an oscillatory shear test was performed at a constant low strain amplitude ($\gamma_0 = 1\%$) with the $\omega$ varied logarithmically from 0.0628 to 62.8 rad/s. Evolution of corresponding moduli probes the linear viscoelasticity of the matured gel. The angular frequency window for such an experiment is decided based on how the experimental (or observation) time $t_{expt}$ for the gels, during gelation rheology at different $\omega$, compares with their $t_{mu}$. The material property, $G'(\omega)$, is constant, i.e., time-independent when $t_{expt} \ll t_{mu}$. However, aging/ thixotropy effects are inherent in $G'(\omega)$ if $t_{expt} \geq t_{mu}$. Subsequently, an oscillatory shear test is performed at a fixed angular frequency ($\omega =$ 6.28 rad/s) while the $\gamma_0$ is varied logarithmically from 0.01 to 1000 %. Evolution of corresponding moduli is examined to understand the nonlinear viscoelasticity of the matured gel. We note that the number of oscillation cycles (typically, 5-10) at each measurement point in the frequency/ amplitude sweep tests is dictated by the default instrument settings. As per standard industrial practice, this ensures that corresponding steady periodic states (i.e., stable signals) are achieved before reporting the rheological properties for most soft materials.

## 3. Results and discussion

### 3.1 USANS

Figure 2 shows $I_{slit}(q)$ measured from un-gelled reconstituted skim milk samples at a CM volume fraction typical of milk concentrations used for gelation (10 % w/w). All curves prior to gelation and the reconstituted skim milk have the same features: a rapid increase in the scattered intensity from high-q to low-q followed by a plateau with decreasing q-values, q < 0.001 $Å^{-1}$, corresponding to the scattering of a compact object of size smaller than 0.1 μm. This length-scale is expected from the casein micelles (CM) in solution[14]. The clear plateau suggests the absence of significant aggregation of the CM in the milk powder reconstituted by the method discussed in Section 2.

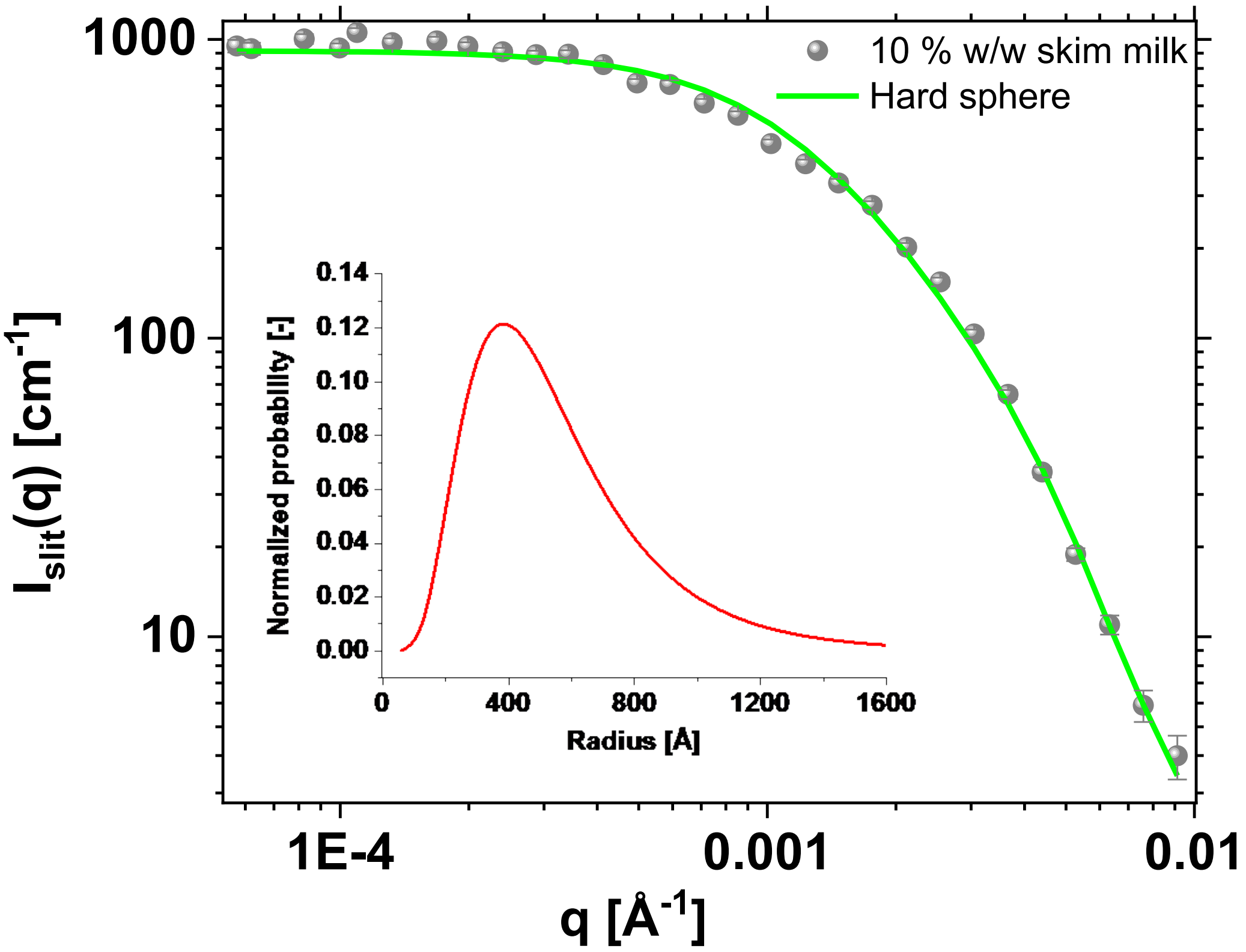

**Figure 2:** Slit smeared scattering cross-section from $D_2O$ reconstituted skim milk powder at a concentration of 10% w/w (grey points), and the best fit obtained from slit smeared version of the model described in Eqn. 1 (green line). The inset shows the radius log normal size distribution corresponding to the hard sphere structure factor used to fit the data.

The analysis of USANS curves from reconstituted milk has been discussed previously.[4, 5, 7, 42] In order to extract important structural parameters from $I_{slit}(q)$ for subsequent use in the analysis of the gel USANS data, the fit here is constrained with reasonable values for: $\phi$ in Eq. 1 is taken equal to the volume fraction of CM, $\phi_{CM}$ , defined by Eqn. 3, reported in Table 3; and the contrast term ($\delta_{CM}$ - $\delta_{solvent}$) calculated from Tables 1 and 2. In this case, we have used only the mean radius (R) and the width of log normal size distribution (i.e. relative standard deviation, PD) of the spheres as the fitting parameters with the structure factor of hard sphere repulsion[40, 49] modified for polydispersity.[40] The best fit is shown along with USANS data for the 10% w/w reconstituted skim milk in Figure 2. The size distribution which was used for the fitting of subsequent data from gelled samples can be seen in the figure's inset. Another structure factor which contains an additional short range attractive term, i.e. the sticky hard sphere[50] was also explored for skim milk by Smith et al.[42, 43] Similar to their work, we observed that our fits with both these structure factors were approximately equivalent and the effect of considering a short-range attraction on the fitting was to alter the average radius and the width of the size distribution used to minimize the difference between the data and the model calculation (not shown here). While it is known that the short range repulsions between CM's are a combination of steric and charged nature of the κ-casein pendants,[25] the physical basis for the short range attractive interaction proposed by Smith et al.[42, 43] is unclear, and therefore will not be utilized further. The volume-fraction–normalized USANS profiles for our reconstituted skim milk reveal that hard sphere structure factor is the most appropriate for our work (see Figure S1 and Section 1 of Supplementary Information).

**Scattering during gelation:** The variation of slit smeared scattering cross-section from reconstituted skim milk samples at 10% w/w during enzymatic and acid coagulation is now discussed. In the early stages of gelation, the curves (not shown here) display similar features to that of reconstituted skim milk of Figure 3. For acid gelation, the first kinetic curve (Figure 3A, red circles) has been measured between 2 and 4 hours after the onset of coagulation, and so on. Likewise, for enzymatic gelation, the first kinetic curve (Figure 3B, red circles) has been measured between 1 and 2 hours after the onset of coagulation, and so on. Each kinetic curve provides time-averaged structural information within the corresponding time window defined by the coarse temporal sampling of $I_{slit}(q)$ enabled by our USANS approach described in section 2.1. The temporal evolution of the scattering curves is faster for enzymatic coagulation (Figure 3B). With progressing time, the scattered intensity in the low-q plateau region ($q < 0.001$ Å$^{-1}$) increases, slowly coming to an equilibrium where the plateau is replaced by a linear slope of approximately -1.3, designated as 'fractal line' in Figure 3. This indicates the onset and growth of fractal-like gel structure during milk coagulation. While we lacked the precise means to locate the gel transition, we observed no clear difference between the scattering from the two gels except for the timescales. It can be observed that within 2-4 hours for acid gel (Figure 3A), but only 1-2 hours for enzymatic gel (Figure 3B), the fractal scaling is present down to 2 x $10^{-4}$ Å$^{-1}$. For longer times, curves rejoin the fractal line at lower and lower q, while the zero-q limit (i.e., forward scattering, $I(0) = \lim_{q\to 0} I(q)$) increases: these two features reveal the growth of cluster size, cluster mass and an increase in the maximum size observable for the fractal aggregates arrangement, namely $\xi$, which is at the limit of the measurable size range for the longest times. That range is limited by the effective q range of the KOOKABURRA instrument, in other words by the fundamental Darwin limit of the width of the rocking curve of the Si (111).[27]

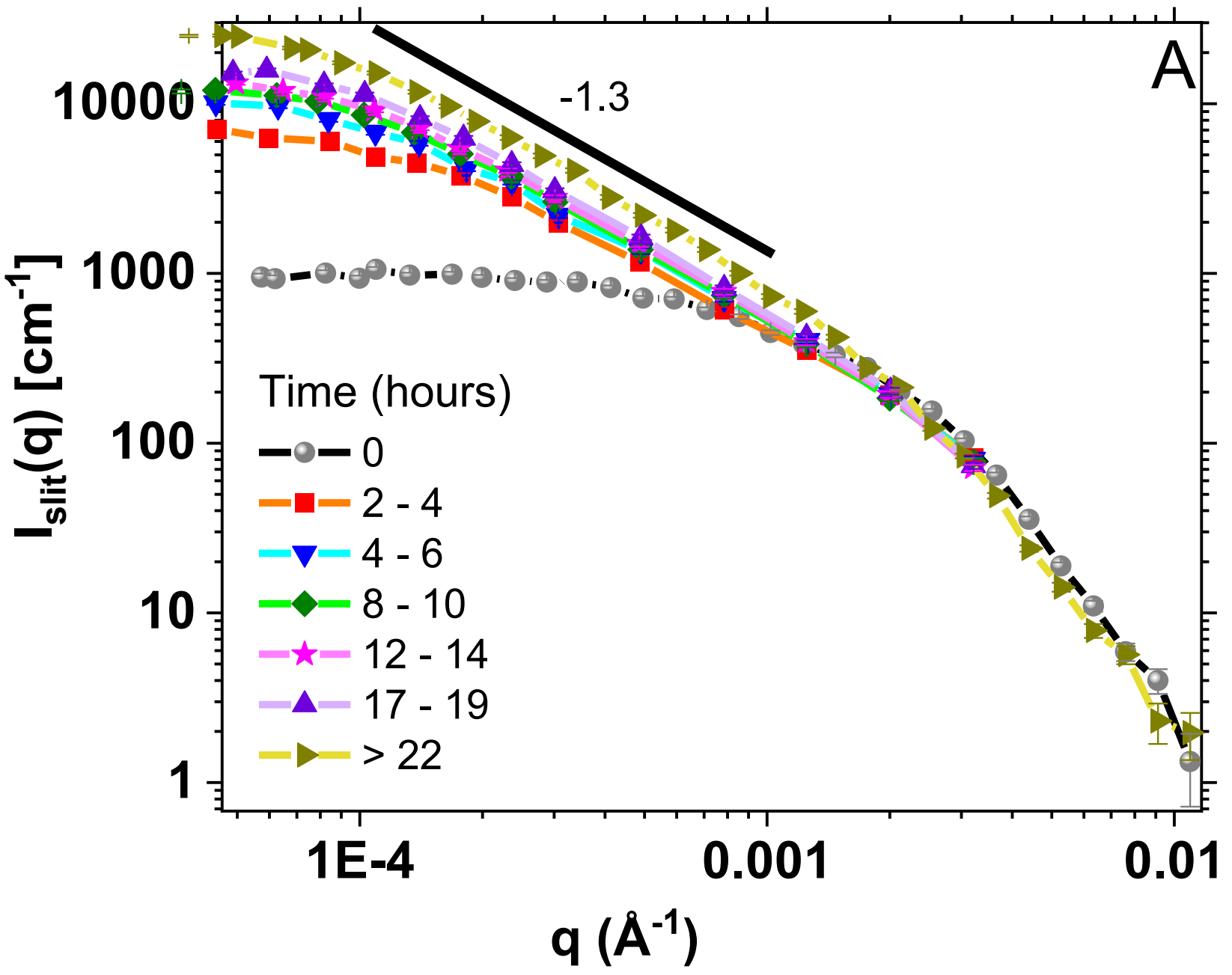

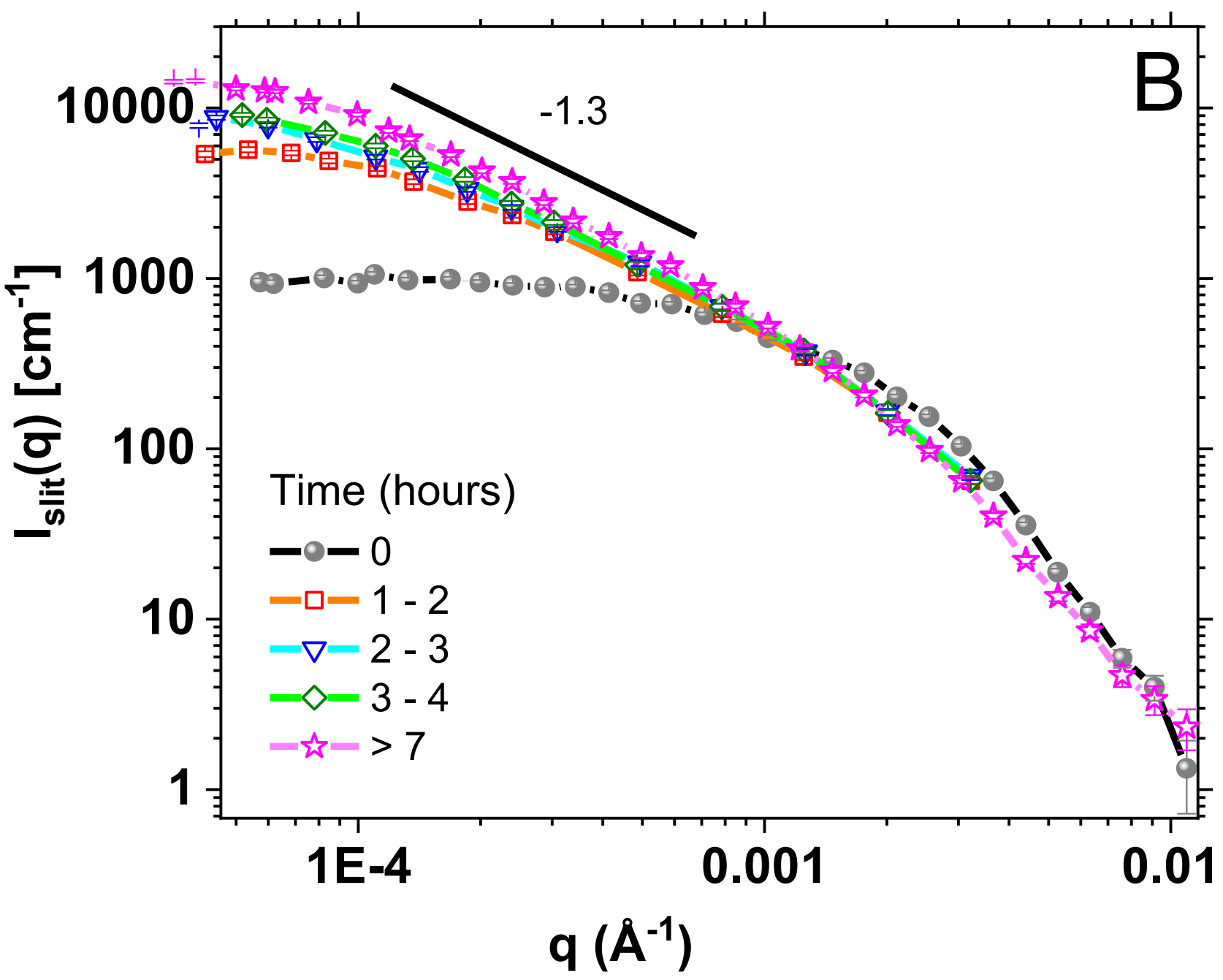

**Figure 3:** Time evolution of the slit smeared scattering cross-section during gelation of $D_2O$ reconstituted skim milk at 10% w/w with acid coagulant **A**; and enzyme, **B**. The fractal line (shown in black) represents the low-q linear slope of the final kinetic curve. The kinetic curves are labelled as per their temporal sampling windows.

We note that the scattered intensity in the high-q region (q > 0.001 Å$^{-1}$) exhibits good superposition at all times for both the acid (Figure 3A) and enzymatic coagulation (Figure 3B). In the early stages of network formation, dominated by smaller sized CM aggregates, the changes in USANS correspond mostly to the mid-q regime, that region which requires longer counting time and exhibits the smallest changes in scattered intensity making USANS ineffective for studying smaller sizes/early stages of aggregation. Rather, it is best suited for studying the growth of clusters and formation of larger fractal networks which have slower/similar-timescale dynamics (minutes to hours) relative to USANS counting times at low-q.

We now discuss the modelling of USANS data from the gels at structural equilibrium, i.e., the final scattering curve in the kinetic series, using the inputs and outputs of modelling $I_{slit}(q)$ from the reconstituted skim milk ($B_{inc}$, $\phi$, ($\delta_{particle} - \delta_{solvent}$), R and the PD). In this case, we will use the fractal structure factor of Chen and Teixeira[51, 52] to describe the spatial density correlations of the fractal network formed by the aggregation of polydisperse spheres (here, CMs) implemented in the program SASVIEW[39] as given by,

$$S(q) = 1 + \frac{D_F \Gamma(D_F - 1)}{\left[1 + \frac{1}{(q\xi)^2}\right]^{\frac{(D_F - 1)}{2}}} \frac{sin[(D_F - 1)\,tan^{-1}(q\xi)]}{qR^{D_F}} \quad (4)$$

where $D_F$ is the fractal dimension of the network, $\Gamma$ is a gamma function and $\xi$ is the correlation length of the network. Convoluting Eqn. 4 with the vertical resolution, we can obtain optimized fits of the equilibrium scattering profiles of enzymatic (Figure 4A) and acid (Figure 4B) coagulated skim milk gels from 10 % w/w reconstituted skim milk. This also corroborates well with literature which hypothesizes dairy gel formation as the assembly of polydisperse CMs into initial aggregates, viz. cylinder-like clusters of CMs, that grow into a three-dimensional fractal network.[6, 7]

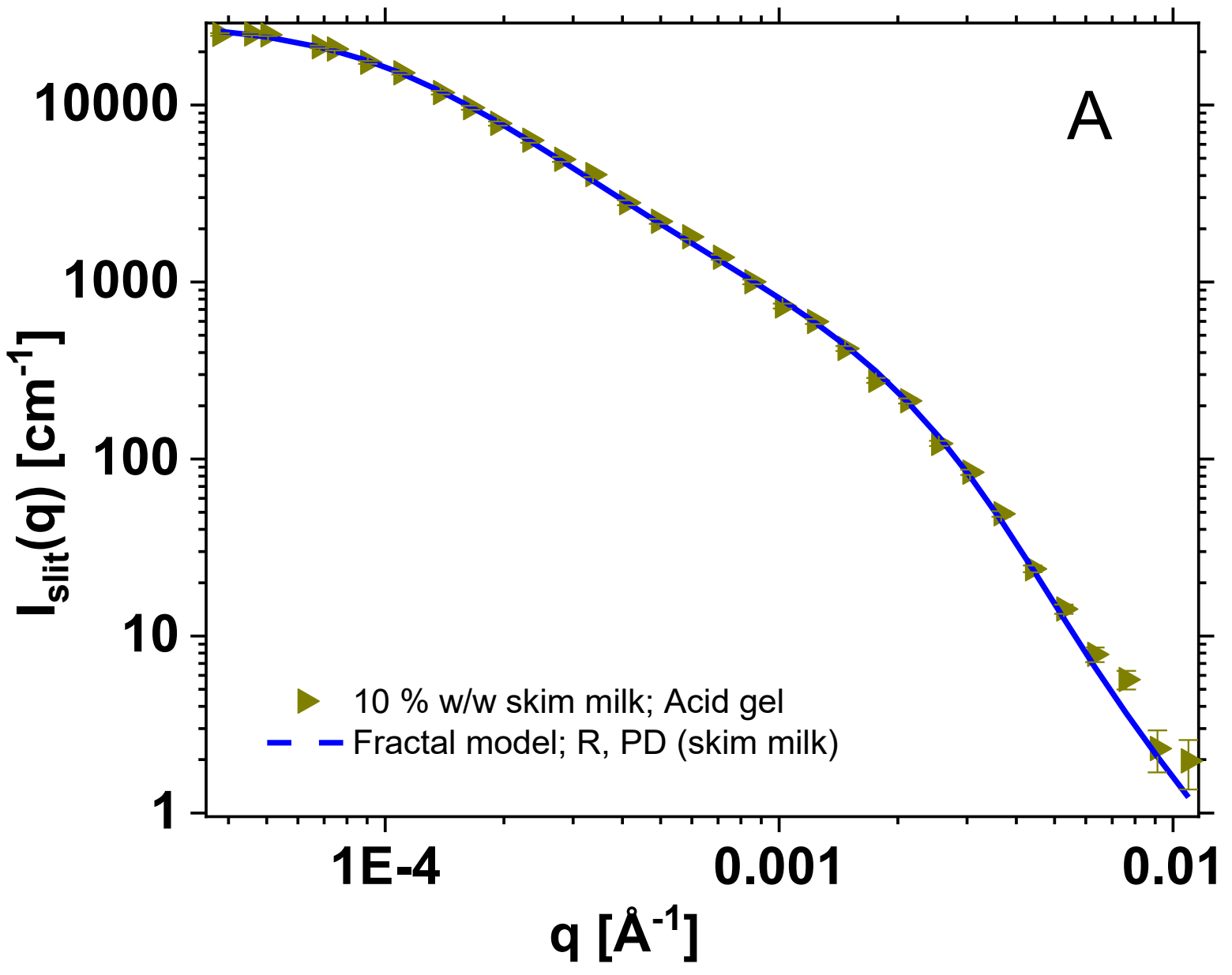

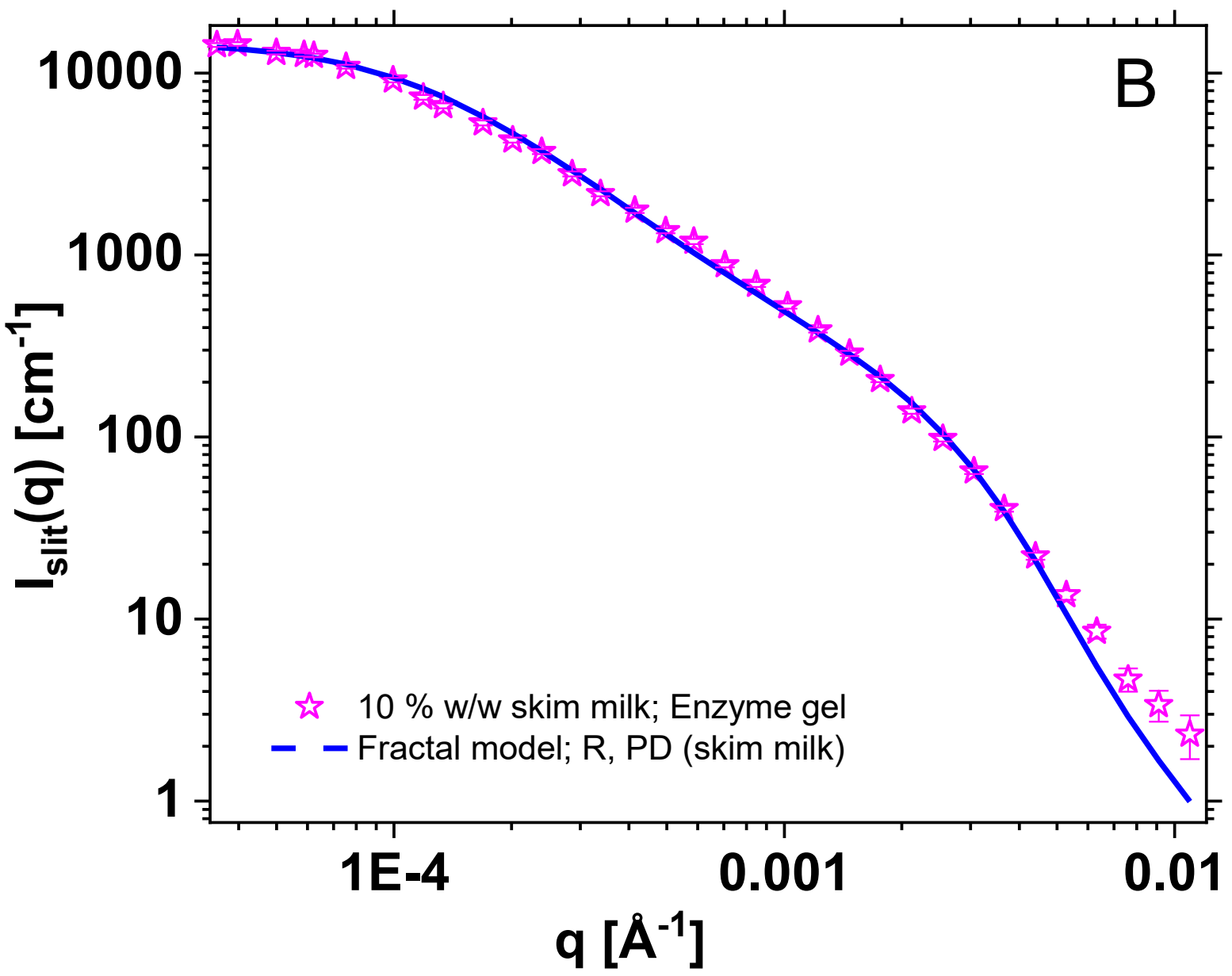

**Figure 4:** The slit smeared scattering cross-section for structurally equilibrated gels, i.e. final kinetic curves (dots) and their best fits obtained from the slit smeared version of Chen & Teixeira fractal model (Eqn.'s 1 & 4) using the computed R and PD of $D_2O$ from the fit (Figure 2) to the scattering of reconstituted skim milk at 10% w/w (Table 3), during acid coagulation, **A**, and enzymatic coagulation, **B**.

The parameters for the fractal network obtained from the USANS fittings of the coagulated gels are presented in Table 3. The fractal dimensions ($D_F$) of the acid and enzyme gels are almost equal, within the error bars, and > 2, indicating that both the coagulation processes follow RLCA-like kinetics, which is consistent with the work of others.[5, 6] Please note that the $D_F$ values are ~ 2.5: the slope for log I(q) versus log q, is larger by 1 with respect to the slope of log $I_{slit}$(q) versus log q.

,

**Table 3:** Parameters obtained from the structural modeling of USANS for reconstituted skim milk, enzyme and acid coagulated gels.

| System | $P(q)$ | $S(q)$ | $R$ (Å) | $D_F$ | $\xi$ (Å) |
|---|---|---|---|---|---|
| Reconstituted skim milk | Spheres with log-normal polydispersity | Hard sphere | 494.86 ± 3.48 | - | - |
| Enzyme gel | Same as above | Fractal | Same as above | 2.503 ± 0.011 | 8101.6 ± 207.39 |
| Acid gel | Same as above | Fractal | Same as above | 2.480 ± 0.009 | 9622.1 ± 270.49 |

The $\xi$ value is of the order of 1μm for both the coagulated gels. Within the error bars $\xi$ for the acid gel is slightly larger than that of the enzyme gel. Translating in real space, this indicates that acid coagulation builds a gel with bigger clusters, larger pores and a more extended network. Enzymatic coagulation leads to a gel with truncated clusters, smaller pores and a finer network. Therefore, both coagulated gels have similar internal mass-scaling and microscopic architecture but slightly differ in the spatial extent of their network structure. Altogether, we can remark that the fractal dimension here does not vary, within the error bar, between 1 or 2 hours of skim milk coagulation and a much longer time. In other words, we do not see further rearrangement at the scale of the network from the perspective of USANS.

### 3.2 Rheology

#### 3.2.1 Cross-point, gel point and critical gel.

Rheologically, when a system undergoes transition in its behavior from liquid-like to solid-like during the process of gelation, the moduli exhibit a gradual change from $G''$ (viscous/ loss modulus) > $G'$ (elastic/ storage modulus) to $G' > G''$. Therefore, a first criterion for observation of gelation should be the crossover in moduli[53]. However, for crosslinking polymers and colloidal gels, the gel point is identified as the onset of a soft-solid with infinite relaxation time, often occurring in the vicinity of the moduli crossover, where it has peculiar viscoelastic properties. Namely, according to the Winter-Chambon criterion,[53,54] applicable for such systems, the gel point is independent of the observation timescale (beyond a minimum value - the local relaxation time characteristic of the mesh size), i.e., the self-similarity of network is spanning over the full volume space of the sample. This state is called a "critical gel", and the corresponding moduli exhibit identical power-law angular frequency ($\omega$) dependence given by,

$$G' = \frac{G''}{\tan\frac{n_c\pi}{2}} = \frac{\pi}{2\Gamma(n_c)\sin\frac{n_c\pi}{2}} S\omega^{n_c} \tag{5}$$

where S is the gel network strength and $\Gamma(n_c)$ is the Euler gamma function of the critical relaxation of the exponent $n_c$. Note that $n_c$, can vary between 0 and 1. For the specific case of $n_c = 0.5$ the gel point coincides with the crossover in moduli (i.e., loss factor, $tan\ \delta = \frac{G''}{G'} = 1$). This is not observed here (see discussion below). There are several values of $n_c$ depending on the type of system, and the mechanism of gel formation.[29,54-57]

In order to detect the crossover point at the appropriate angular frequencies, Figure 5A shows the variation of $\tan\delta(\omega)$ with time during enzymatic coagulation of the representative 10 % w/w reconstituted skim milk at three different angular frequencies $\omega_1$, $\omega_2$, $\omega_3$. It can be observed that, at initial times, $\tan\delta(\omega)$ values are noisy due to sol-like state of the system for which $G'$ is negligible. As gelation proceeds a steady decrease of $\tan\delta(\omega)$ is prevalent at short times (t < 3000 s) eventually stabilizing to a constant low value at longer times. This indicates that the enzymatic coagulation has started and the range of 3.14 rad/s – 31.4 rad/s is appropriate to detect it. At the lowest angular frequency ($\omega_1$ = 3.14 rad/s, slowest motions), the system shows lower $\tan\delta$ values in the sol-state indicating that the oscillatory shear is slow enough for the underlying structure to stretch/relax. At the highest angular frequency ($\omega_3$ = 31.4 rad/s), the system shows higher tan $\delta$ values in the sol-state implying that the shear is too fast for such deformation, i.e., $G''(\omega_3)$ is relatively higher than $G''(\omega_1)$. Thus, we choose $\omega_2 = 6.28$ rad/s as the reference angular frequency because it lies in the middle of the measured angular frequency window, where tan $\delta$ provides a representative balance of viscous and elastic contributions. We will later use this reference angular frequency for monitoring at all skim milk concentrations, as shown in Figure 5B.

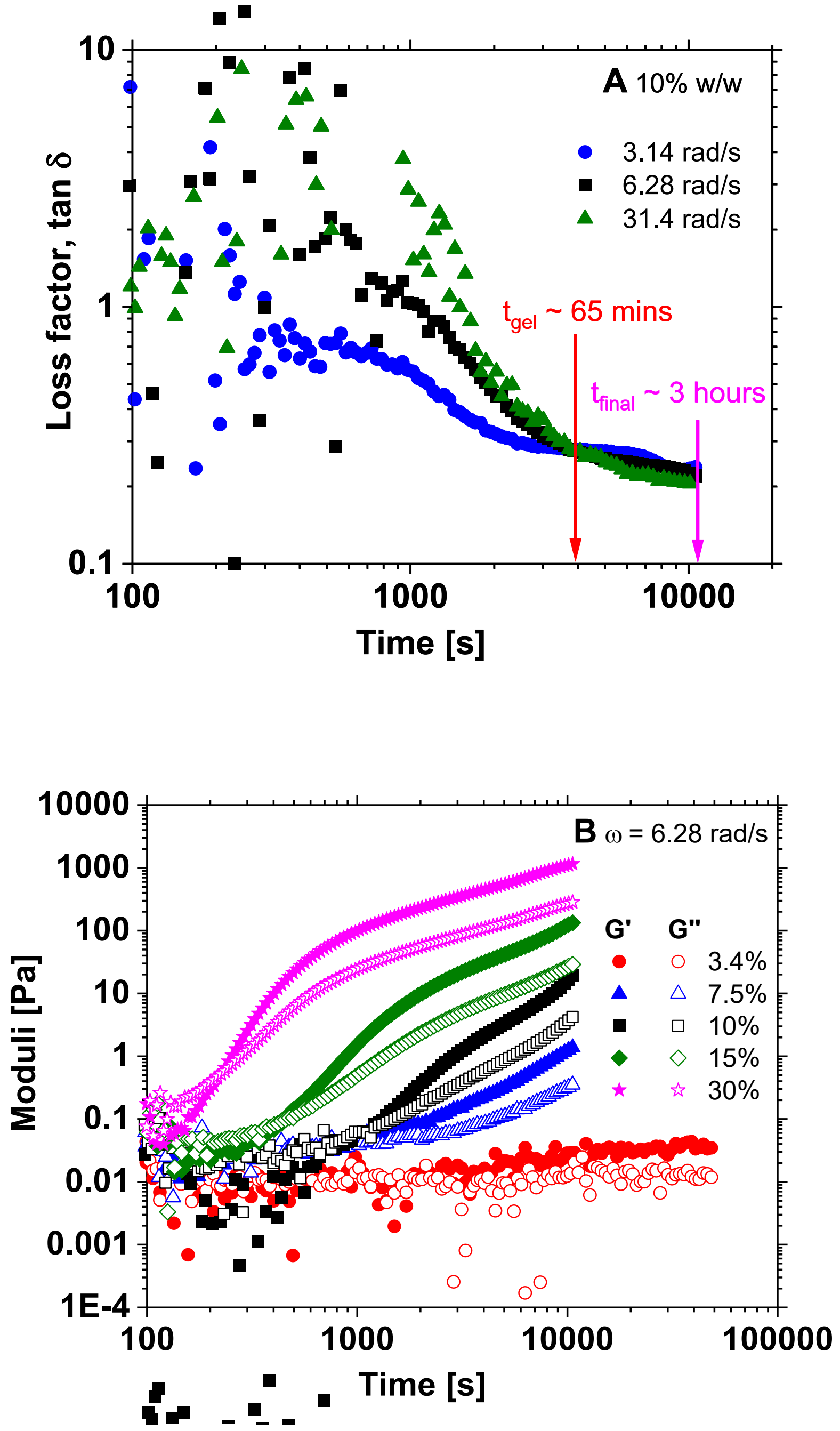

**Figure 5.** Oscillatory time sweep response from enzymatic coagulation at $\gamma_0 = 1\%$. Loss factor, $\tan\delta$, as a function of time for gelation of 10% w/w reconstituted skim milk at $\omega =$ 3.14, 6.28, 31.4 rad/s, **A**; Moduli $G'$ and $G''$ as a function of time during gelation of 3.4, 7.5, 10, 15 and 30 % w/w reconstituted skim milk at the reference $\omega = 6.28$ rad/s, **B**.

For the 10 % w/w reconstituted skim milk during enzymatic coagulation, the $\tan\delta$ curves at the three angular frequencies of 3.14, 6.28, and 31.4 rad/s coincide at the time $t_{gel}$ ~ 65 minutes (Figure 5A). If $\tan\delta(\omega)$ is constant, the curves $G'(\omega)$ and $G''(\omega)$ are parallel, and the corresponding time $t_{gel}$ can be identified as the critical gel point, according to the Winter-Chambon criterion recalled above. Note that $t_{gel}$ here is longer than the time at which $G'(t)$ crosses $G''(t)$, which depends on angular frequency. We now use Eqn. 5 and obtain $n_{c-gp} = 0.17$ from the intersection value of $\tan\delta$ curves (Figure 5A), $\tan\delta_{gp}$, here equal to 0.27. The Muthukumar relation, originally derived for crosslinking polymers,[29] combines the dynamics of polymer fractals – assuming that the dynamics of CM chains are same as for polymers, with percolation network theories to obtain the fractal dimension $d_f$ of the network in the vicinity of sol-gel transition, as given by:

$$n = \frac{d\left(d + 2 - 2d_f\right)}{2\left(d + 2 - d_f\right)} \tag{6}$$

where $d$ (=3) is the Euclidean dimension. Accepting this relation for our critical gel point (i.e., $n = n_{c-gp}$), we obtain $d_{f-gp} = 2.35$ agreeing with our proposition that the first step of enzymatic coagulation is an RLCA-like process, and consistent with the value obtained by USANS starting to be visible around 30 min but fully established at longer time 240 min. Although our system is not made of linear polymer, it appears that Eqn. 6 can be applied, as assumed formerly for colloidal gelation.[28, 58]

### 3.2.2 Time evolution beyond gel point:

#### 3.2.2.1 Evolution of tan δ

We now analyze the tan δ variation including longer times than $t_{gel}$, at the reference angular frequency of 6.28 rad/s, during enzymatic coagulation of reconstituted skim milk at different concentrations (Figure 6A below).

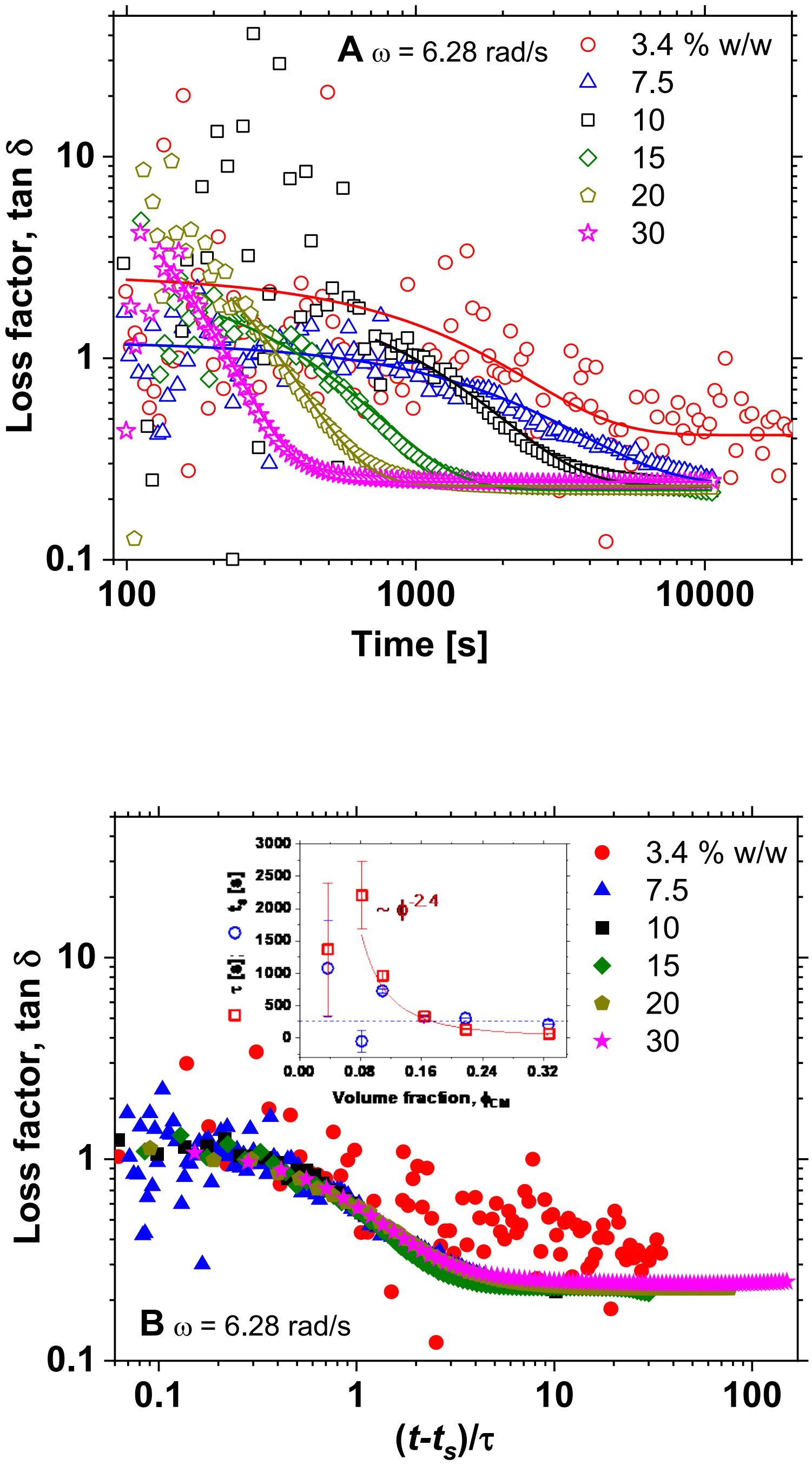

**Figure 6.** Temporal evolution of the loss factor tan δ, at the reference $\omega$ = 6.28 rad/s, during enzymatic coagulation of reconstituted skim milk at 3.4, 7.5, 10, 15, 20 and 30 % w/w concentrations (dots), corresponding to the moduli in Figure 5B. Curves are the best fit of this data to the exponential decay model (Eqn. 7), **A**; The same tan $\delta$ curves displayed as function

of the dimensionless time $\tilde{t} = (t - t_s)/\tau$. Inset: characteristic time $\tau$ (red squares) and empirical shift factor $t_s$ (blue circles) as function of the volume fraction $\phi_{CM}$ at each skim milk concentration. Red curve is the best power law fit of $\tau$ vs. $\phi_{CM}$. Blue line guides the eye along the variation of $t_s$ vs. $\phi_{CM}$, **B**.

In Fig. 6A, an exponential-like monotonic decay becomes evident for all concentrations, which is described by the following equation:

$$\tan\delta(t) = e^{-\left[\frac{t-t_s}{\tau}\right]} + k \tag{7}$$

where $t_s$ is the concentration-dependent empirical time shift factor, $\tau$ the characteristic time and $k$ plateau value of $\tan\delta(t)$ at long time, respectively. Then, as shown in Figure 6B, the tan δ curves at different skim milk concentrations appear to nearly collapse onto a single master curve when plotted as a function of the dimensionless time $\tilde{t} = (t - t_s)/\tau$,. This universal law of time-concentration superposition reveals kinetic similarity of the underlying gelation process. This indicates that the enzymatic coagulation of skim milk follows a global kinetic pathway governed by CM aggregation, fractal clustering and restructuring - although it briefly passes through a Winter Chambon-like quasi-critical state. Likewise, we report the reconstituted skim milk at different concentrations to exhibit kinetic similarity during acid coagulation as well (Figure S3, Supplementary Information). Such behavior is reminiscent of gelation in biopolymers, proteins and colloidal particles wherein percolation-type critical scaling exists only around the gel point while the network goes on building via broader non-critical aggregation kinetics, in two step aging process.[55, 56, 58, 59] In particular, this has been observed in the reconstituted milk gelation study by Bauland et al.[55] Looking further at the parameters obtained from the exponential fits as per Eqn. 7, $t_s$ is the point where tan δ begins to sharply decrease indicative of the onset of percolation in the critical gel region, whereas $\tau$ reflects the steepness of tan δ decay which represents the inverse of the overall rate/kinetics sign of gelation.

For both enzymatic and acid coagulation in our work, $t_s$ remains almost independent of the skim milk concentration (or $\phi_{CM}$) while $\tau$ shows a strong power law variation with $\phi_{CM}$ (insets of Figure 6B, and Figure S3C in supplementary information). The behavior is $\tau \sim 3.65\ \phi_{CM}^{-2.4}$ for enzymatic coagulation and $\tau \sim 0.07\ \phi_{CM}^{-3.4}$ for acid coagulation. From the corresponding $\tau - \phi_{CM}$ power-law exponents, we deduce that the gelation kinetics accelerate as the skim milk concentration increases, more strongly for the case of acid coagulation than that of enzymatic coagulation. From the $\tau$ obtained at the $\phi_{CM}$ specific to 10 % w/w skim milk concentration, we can further measure slower kinetics (i.e., lower gelation rate due to higher value of $\tau$ at larger $\phi_{CM}$) for enzymatic coagulation than that of acid coagulation, contrary to the USANS observation. This suggests that the macroscopic elasticity percolation and the microscopic structure formation are not completely correlated during skim milk coagulation possibly because network rigidity involves additional mechanisms probably at different length-scales than that probed by USANS.

**3.2.2.2 Evolution of fractal dimensions.**

We recall that in this work the crossover in moduli at the reference angular frequency for the representative 10% w/w reconstituted skim (Figure 5B) occurs at $t_{cross}$(6.28 rad/s), ~ 30 minutes after the initiation of enzymatic coagulation, and well before the critical gel point ($t_{gel}$ ~ 65 minutes). Using again Eqn. 5 to obtain $n_{c-cross} = 0.5$ from $\tan \delta_{cross}$, here equal to 1, the Muthukumar relation[29] (Eqn. 6) provides $d_{f-cross} = 2$ at $t_{cross}$(6.28 rad/s). Between this occurrence at $t_{cross}$(6.28 rad/s) and $t_{gel}$, the elastic modulus increases continuously to higher values at all $\omega$. This implies that the system goes through gelation in a multi-stage fashion. Firstly initiation of network occurs wherein rapid cluster coarsening leads to the emergence of a cluster spanning over the whole system (i.e. percolation). Secondly, the network is consolidated, probably by cluster tightening and/or densification). Finally, a late-stage network stiffening due to local restructuring (i.e. arrested coarsening of clusters) can also be inferred from the slightly differing values of $d_{f-cross}$ (~2) at $t_{cross}$(6.28 rad/s), $d_{f-gp}$ at $t_{gel}$(~2.35) and $d_{f-USANS}$ (~2.5) obtained from USANS at longer times (e.g. 2-3 hours). This indicates that the system has several successive evolutions, a signature typical of aging kinetics. There have been recent investigations of the instantaneous relaxation spectrum by Bauland et al[58] . Using this technique, they accurately examined the viscoelasticity throughout gelation and reported a two steps aging during the enzymatic coagulation of skim milk. . After the percolation of the network, the authors made also use of confocal microscopy[58] indicating rapid coarsening of the underlying microstructure followed by an arrest of any dynamics. The aging regime starts around the $t_{gel}$ , and evolves at a much slower pace via a purely contact-driven mechanism. They report no further microscale rearrangements at this stage from confocal microscopy[58]. So the fractal scaling remains the same in that size range, across the critical gel and post-gelation states despite the subtle temporal changes in moduli.

In the present study, the maximum observed value of $t$ should be far larger than $t_{gel}$ since we wish to consider the properties of fully developed network structure. The temporal evolution of $\tan\delta$ for the representative 10 % w/w reconstituted skim milk reaches a plateau within ~ 4000s for enzymatic coagulation (Figure 6A, black dots), and ~2000s for acid coagulation (Figure S3B in supplementary information, black dots). The rheology is continued further than these values, until 3 hours to obtain a mature gel state. For this time of ~3 hours we have result from our USANS studies scattering shows fractal slope over the largest part of the q-range, and stabilizes for both enzymatic and acid coagulation (as discussed earlier, Figure 3) when reaching 4 h.

For rheology, the time evolution depends on concentration. Figure 5B compares the evolution of moduli during enzymatic coagulation of reconstituted skim milk at different concentrations.

We observe that the crossover time between $G'$ and $G''$ (at 6.28 rad/s) decreases, i.e., a faster arrival of the mature gel-state occurs, with increasing skim milk concentrations (i.e. increasing $\phi_{CM}$). The corresponding moduli also increase, indicating that stronger enzymatic gels are formed at higher $\phi_{CM}$, as expected. Similar observations can be derived from the acid coagulation of reconstituted skim milk at concentrations up to 10 % w/w – i.e. $\phi_{CM}$ = 0.109 (Figure S3A, supplementary information). The deviation at 15 % w/w is merely an effect of low acid coagulant availability for higher skim milk content, as is known in literature.[30, 60]

In another recent work from Hemar's group[4], time-resolved SANS/ USANS and gelation rheology were utilized to study the acid coagulation of skim milk (at ~ 4 wt. % protein content) from various sources. The authors were interested in nano-structural evolution of colloidal calcium phosphate (CCP) within the CMs, alongside CM aggregation and network formation. It can be seen from their data that the time taken for the fractal power-law exponent in USANS to reach a steady value coincides with the time of rise in $G'$ during rheology.[4] This indicates that their USANS measurement reflects the onset of the mature gel state, as in our work.

### 3.2.3 Oscillatory shear: frequency and amplitude dependence in mature gels

**Note on mutation times.** In order to ensure that the longer time rheological characterization of the mature gels in our work is not perturbed by aging effects, we calculated mutation times $t_{mu}$ to compare with the duration of the measurements. The $t_{mu}$ values are found to range from ~100s around the gel point to >1000s in the mature gel (refer Section 4, Figure S4). Hence ranges for the subsequent frequency sweep measurements are chosen such that the measurement time (i.e., inverse of frequency, $f = \omega/2\pi$) is shorter than $t_{mu}$ at least by a factor of 10, in all cases.[48, 61]

### 3.2.3.1 Effect of angular frequency:

**Enzyme coagulated gels.** The mature gels obtained from gelation rheology of reconstituted skim milk at different concentrations during enzymatic coagulation, carried out for a duration of 3 hours, are immediately subjected to frequency sweep rheology (Section 2.2, B). The corresponding response at representative skim milk concentrations of 3.4, 10, and 30 % w/w are shown in Figure 7A.

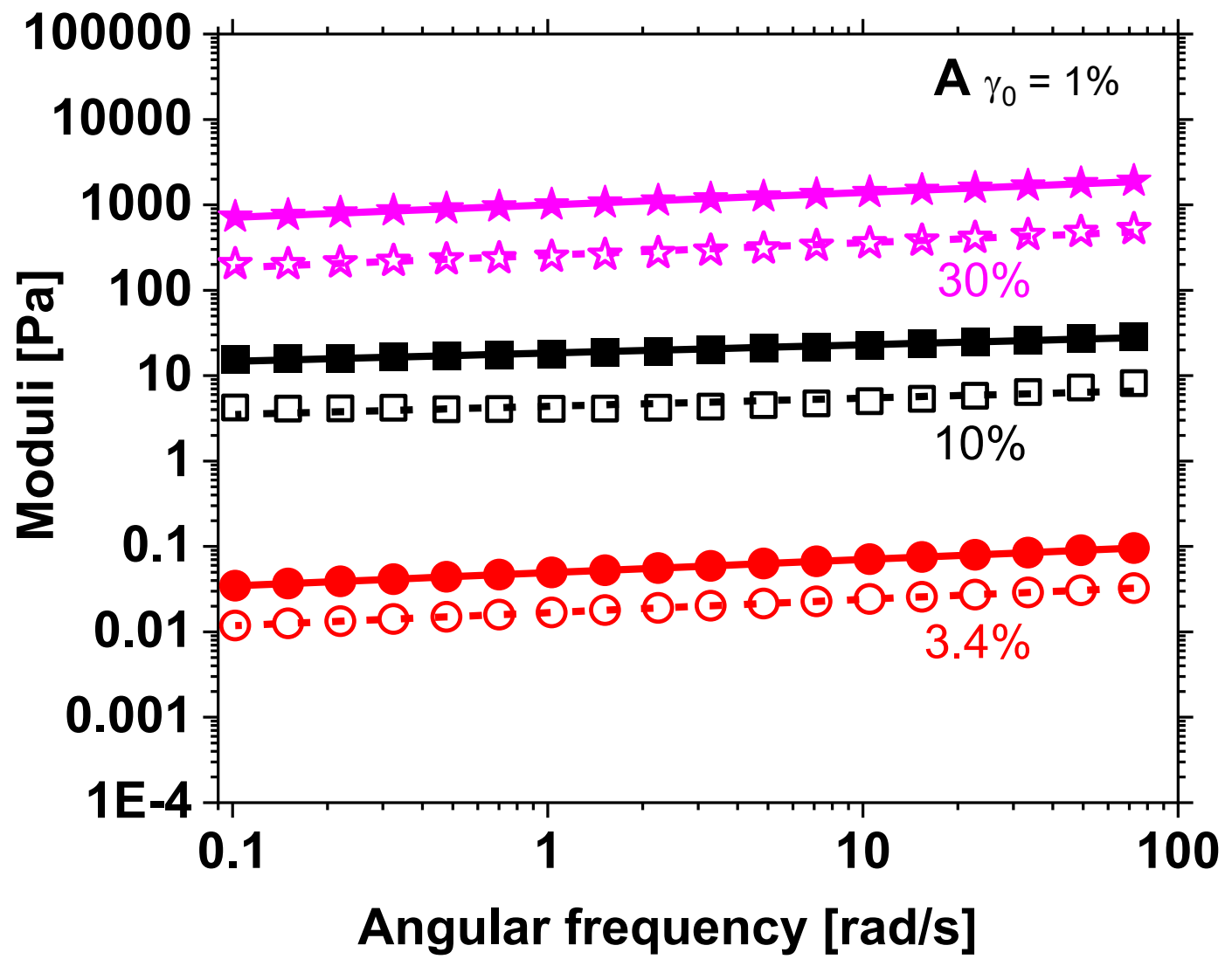

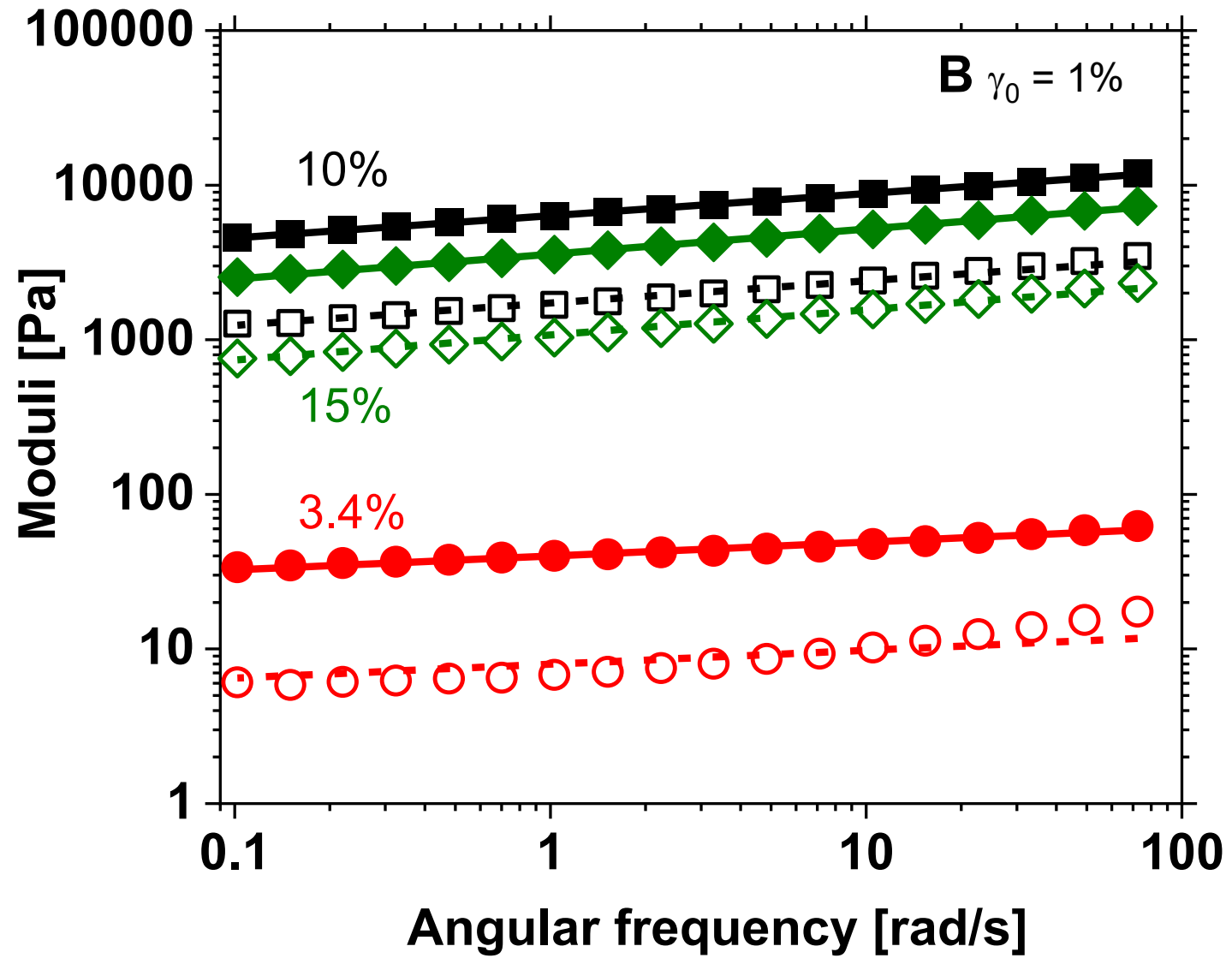

**Figure 7.** Moduli $G'$ and $G''$as a function of angular frequency at $\gamma_0 = 1$ % during oscillatory shear of mature gels from reconstituted skim milk at different concentrations obtained via:

Enzymatic coagulation (up to 3 hours), **A**; Acid coagulation (up to 3 hours), **B**. The solid lines represent fits according to the modified weak gel model (Eqn. 9).

In all cases, it is observed that the moduli, $G'$ and $G''$, exhibit a weak angular frequency dependence which is common to most mature colloidal gels[20]. Such behavior are effectively described by so-called fractional constitutive models[62]. We use the simplest form here which involves a fractional viscoelastic element that interpolates between an elastic spring and a viscous dashpot (noted "spring-pot"), given by:

$$\sigma = G_0 \frac{d^n \gamma}{dt^n} \tag{7}$$

where $\sigma$, $\gamma$, and $G_0$ represent the stress, strain, and a quasi-property with dimensions $Pa.s^n$ representing the fractal gel strength, respectively: the 'spring' and 'dashpot' are mechanical analogues of a purely elastic solid which stores energy and a purely viscous fluid that dissipates energy, respectively. Then the dimensionless exponent $n$ can vary between 0 (spring) and 1 (dashpot). For oscillatory shear flow conditions, Eqn. 7 reduces to the weak gel model[63], given below.

$$G^*_{wg}(\omega) = G_0(i\omega)^n = G_0\omega^n \cos\frac{n\pi}{2} + iG_0\omega^n \sin\frac{n\pi}{2} \tag{8}$$

Although Eqn 8. resembles the viscoelastic scaling (Eqn. 5) at the gel point ("critical gel") it was developed using a different approach, for food systems, with a finite relaxation time. Systems which have a three-dimensional network, similar to the gels far beyond their gel point, have high values of moduli but are also capable of flowing[64]. Thus, the theory of cooperative flow was employed where network strands are assumed to be composed of interacting 'flow units'.[65] The network strands break during shear, but not the flow units, leading to a power-law behavior. The corresponding exponent is $n = 1/z$, where $z$ is the coordination number representing the number of flow units connected together in the system.[63] A modified version of this model is also available where an additional phase shift term, $\varphi_{shift}$, is introduced such that $\varphi = \frac{n\pi}{2} + \varphi_{shift}$ in order to account for the difference in network elasticity with the one predicted by the fractional model[66], as follows:

$$G^*_{mwg}(\omega) = G_0(i\omega)^n e^{i\varphi_{shift}} = G_0\omega^n \cos\varphi + iG_0\omega^n \sin\varphi \qquad (9)$$

It is observed that the modified weak gel model fits, shown in Figure 7A, are in good agreement with the experimental moduli of our mature enzyme gels, where $\varphi_{shift} \in (0.01, 0.1)$, giving $n = 0.14 \pm 0.03$, and the $G_0$ values are reported in Figure 9A. From the obtained values of $n$, and using again the Muthukumar relation[29] (Eqn. 6), although in a slightly different work-frame, provides $d_f \sim 2.38$. This is similar to the $d_f$ obtained by Bauland et al[58] for the mature gels from enzymatic coagulation of their skim milk, although the fractional model used by them, and the relevant exponents, differs slightly from our work. Apart from that, we note that the magnitude of both moduli during frequency sweep of our mature enzyme gels increases with skim milk concentration. This is discussed in more detail later.

**Acid coagulated gels.** The frequency sweep response of mature gels obtained from gelation rheology of reconstituted skim milk at different concentrations during acid coagulation, carried out for a duration of 3 hours, are shown in Figure 7B. It is observed that the moduli, $G'$ and $G''$, exhibit here also a weak angular frequency dependence, and their magnitude increases with the skim milk concentration up to 10 % w/w, but this time, for a higher value, 15% w/w, a decrease with concentration occurs. Since no macroscopic syneresis was evident during the experiment, we ascertain that this decrease is merely an effect of low acid coagulant availability for higher skim milk content, as is known in literature [30, 60]. Thus, we only consider mature acid gels up to 10 % w/w skim milk concentration for further analysis. The modified weak gel model (Eqn. 9) is used again, and it perfectly captures the power law behavior of the moduli, where $\varphi_{shift} \in (0.01, 0.1)$, giving $n = 0.12 \pm 0.03$, and the $G_0$ values are reported in Figure 9B. From the obtained values of $n$, and using again the Muthukumar relation[29] (Eqn. 6) provides $d_f \sim 2.39$, similar to that of our mature enzyme gels. We recall that the Muthukumar relation[29] is derived from percolation theories strictly valid for the critical gel region, and its applicability above the sol-gel transition is thus debatable[29]. Therefore, we consider it here with caution and follow a different approach to fractal scaling analysis, discussed later in this section.

### 3.2.3.2 Effect of shear strain:

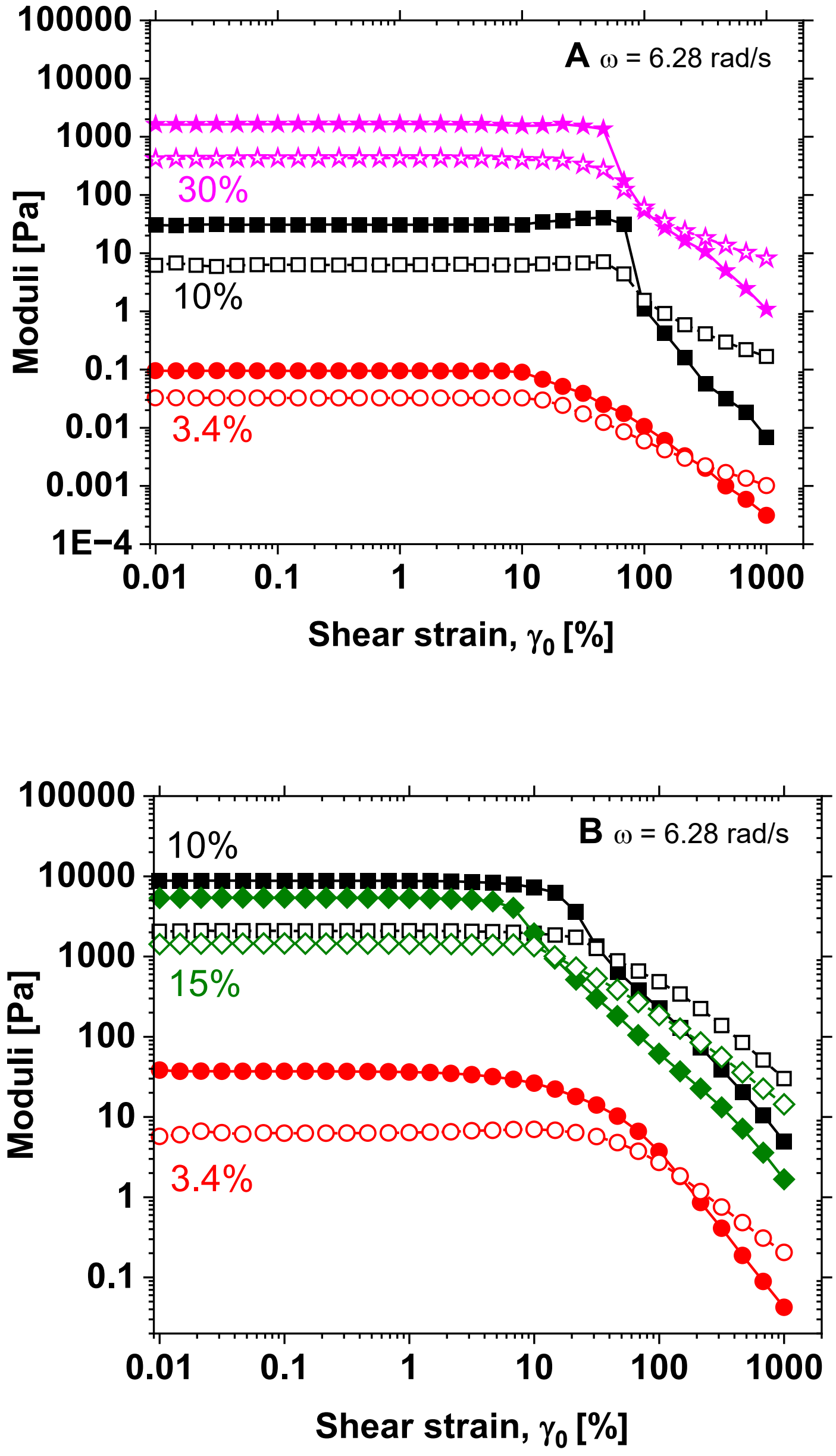

**Figure 8.** Moduli $G'$(closed symbols) and $G''$(open symbols) as a function of the oscillatory shear strain amplitude at $\omega = 6.28$ rad/s, after the frequency sweep rheology, of mature gels from reconstituted skim milk at different concentrations obtained via: Enzymatic coagulation (up to 3 hours), **A**; Acid coagulation (up to 3 hours), **B**. The solid lines are drawn to guide the eye.

**Enzyme coagulated gels.** The mature gels of reconstituted skim milk at different concentrations at the end of frequency sweep rheology during enzymatic coagulation (up to 3 hours), are subjected to strain amplitude sweeps. The corresponding response at representative concentrations of 3.4, 10, and 30 % w/w are shown in Figure 8A. In all cases, it is observed that the moduli are constant with $G' > G''$ (i.e., viscoelastic solid-like) up to a critical strain, $\gamma_{0C}$, beyond which a transition to $G'' > G'$ (i.e., viscoelastic liquid-like) occurs with both the moduli decreasing as a function of $\gamma_0$ (> $\gamma_{0C}$). Very importantly, note that **there is no complete macroscopic breakage of the gel during the test.** A very weak overshoot in $G''$ and $G'$ is also noticed around $\gamma_{0C}$, particularly in the case of 10 % w/w skim milk concentration.

Such a trend in $G'$ ($\gamma_0$) and $G''(\gamma_0)$ is termed as nonlinear Type III behavior according to large amplitude oscillatory shear (LAOS) rheology literature.[67] This is attributed to shear-induced structural changes, such as inter-floc rearrangements which are common to most colloidal/ protein gels.[20,67] Concerning the observed weak $G'$ overshoot, we note that it is also reported by Bauland et al[58] in the strain sweeps of their enzyme coagulated skim milk gels at different aging times. This overshoot indicates an accumulation of elastic stresses ($G'$ rise) with increasing $\gamma_0$, followed by viscous dissipation, i.e., the stress distributes evenly across the network before its breakdown ($G''$ overshoot and $G' > G''$ ). This is a typical signature of gradual plastic-like yielding in the system.[67] Apart from that, we note that the magnitude of both moduli during amplitude sweep of our mature enzyme gels increases with skim milk concentration. This is discussed in more detail later.

**Acid coagulated gels.** The strain amplitude sweep response after frequency sweep rheology of mature gels was obtained from gelation rheology of reconstituted skim milk at different concentrations during acid coagulation (up to 3 hours, shown in Figure 8B). It is observed that while the magnitude of both moduli increases with skim milk concentration up to 10% w/w at all $\gamma_0$ as for the enzyme coagulated gels, the evolution for acid coagulated gels show similarities and differences. It is similar in the viscoelastic solid-like to liquid-like transition with increasing $\gamma_0$, arising from shear-induced structural changes, as described earlier, which are common to most colloidal/ protein gels.[20,67] It is different in that the mature acid gels show no clear overshoot in moduli around the critical strain, $\gamma_{0C}$, indicating that the yielding here is rather sudden with a little/ no plastic-like flow before breakdown.[67] Also, the evolution is much smoother. We can hypothesize that the distinct mechanical behaviour of our two mature gel types is a direct manifestation of their different underlying network structure.

**3.2.3.3 Dependence upon concentration.** We will now quantify the effect of skim milk concentration (or CM volume fraction, $\phi_{CM}$) on the elasticity of our mature gels. The value of $G'(\omega)$ at $\omega = 6.28$ rad/s and $G'(\gamma_0)$ at the low strain amplitude of $\gamma_0 = 1\%$ are designated as the linear moduli from frequency sweep rheology ($G'_{FS}$) and amplitude sweep rheology ($G'_{AS}$), respectively. The $G_0$ values are obtained from modified weak gel model (Eqn. 9) fitting of the frequency sweep data. The strain amplitude at which the moduli crossover and transition from viscoelastic solid-like to liquid-like response in amplitude sweep data is identified as the critical strain ($\gamma_{0C}$). The corresponding stress value is referred to as the critical stress ($\sigma_{0C}$). Together, the couple ($\gamma_{0C}$, $\sigma_{0C}$), is called the linear viscoelastic limit (LVE) beyond which the system exhibits yielding and nonlinear viscoelastic response, according to the LAOS rheology literature.[67]

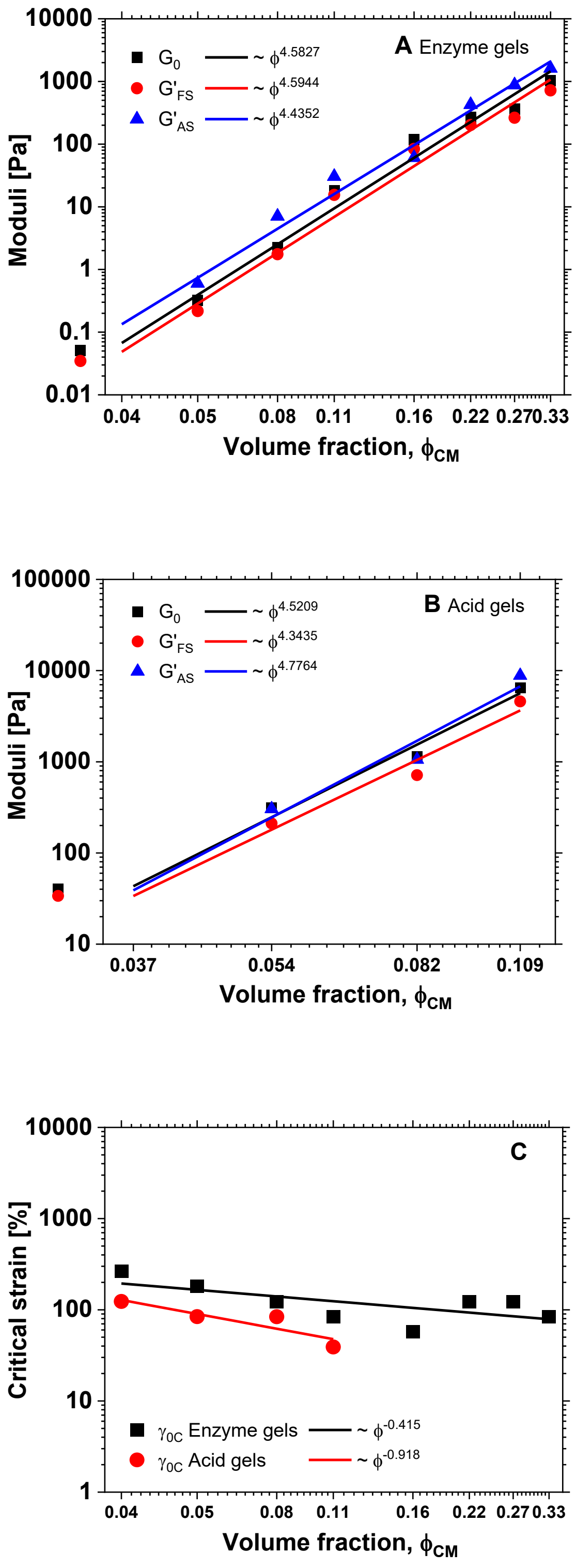
A Enzyme gels
G0 ~ ϕ4.5827
G'FS ~ ϕ4.5944
G'AS ~ ϕ4.4352
Moduli [Pa]
10000
1000
100
10
1
0.1
0.01
0.04
0.05
0.08
0.11
0.16
0.22
0.27
0.33
Volume fraction, ϕCM
B Acid gels
G0 ~ ϕ4.5209
G'FS ~ ϕ4.3435
G'AS ~ ϕ4.7764
Moduli [Pa]
100000
10000
1000
100
10
0.037
0.054
0.082
0.109
Volume fraction, ϕCM
C
γ0C Enzyme gels ~ ϕ-0.415
γ0C Acid gels ~ ϕ-0.918
Critical strain [%]
10000
1000
100
10
1
0.04
0.05
0.08
0.11
0.16
0.22
0.27
0.33
Volume fraction, ϕCM

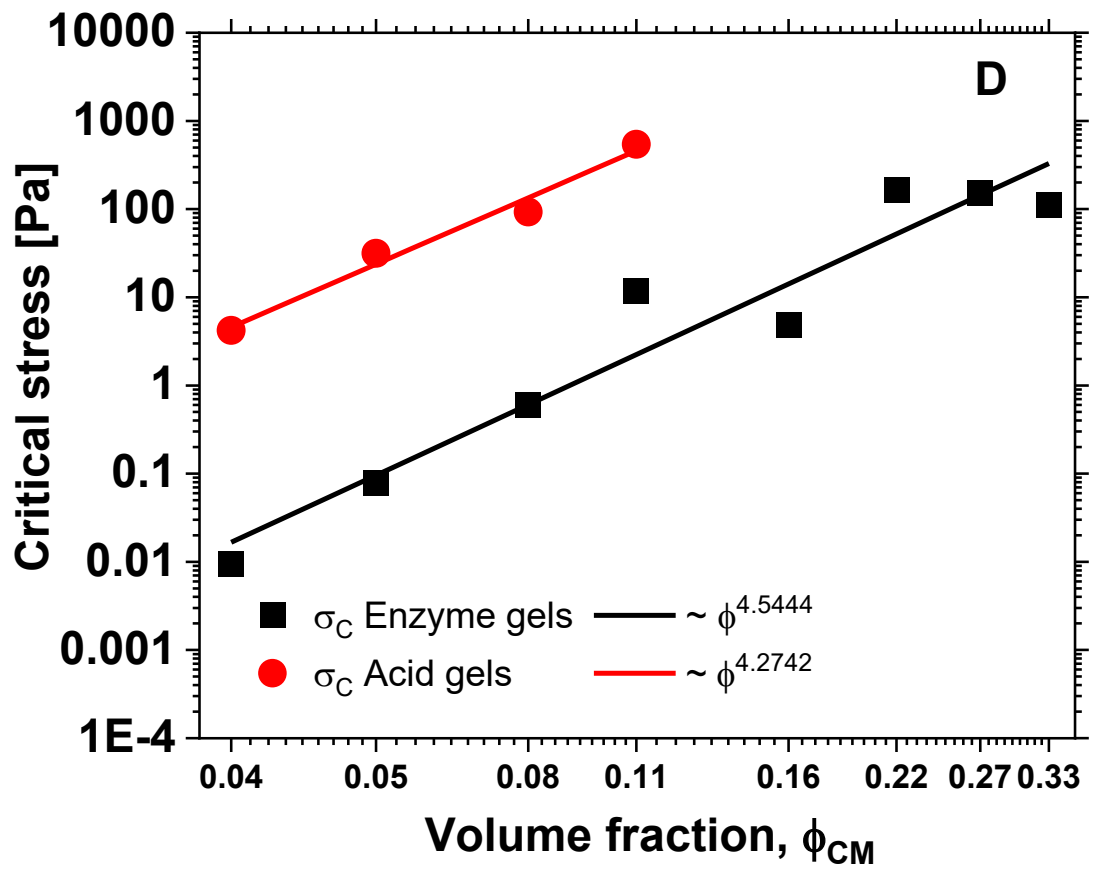

**Figure 9.** The linear moduli (at $\omega$ = 6.28 rad/s) from frequency sweep ($G'_{FS}$), the $G_0$ from modified weak gel model (Eqn. 9) fits of the frequency sweep data, and the linear moduli (at $\gamma_0$ = 1%) from amplitude sweep ($G'_{AS}$) as a function of the CM volume fraction ($\phi_{CM}$) corresponding to the reconstituted skim milk concentrations for matured gels formed via: Enzymatic coagulation, **A**; Acid coagulation, **B**. The variation with $\phi_{CM}$ of critical strain $\gamma_{0C}$, **C**; and corresponding critical stress $\sigma_{0C}$, **D**, obtained from the amplitude sweep of mature gels formed via enzymatic and acid coagulation of reconstituted skim milk at different concentrations. The solid lines represent power-law fits to the experimental data.

The variation of $G_0$, $G'_{FS}$, and $G'_{AS}$ with $\phi_{CM}$ for the mature gels from enzymatic coagulation of reconstituted skim milk is shown in Figure 9A. The corresponding trend for the mature gels from acid coagulation are shown in Figure 9B. It is clearly observed that the linear moduli and $G_0$ for both the mature gels increase with $\phi_{CM}$ following a power-law relationship. Figures 9C and 9D show the variation of $\gamma_{0C}$ and $\sigma_{0C}$ with $\phi_{CM}$ respectively, for the matured gels from enzymatic and acid coagulation of reconstituted skim milk. It is observed that for both enzymatic and acid mature gels, $\gamma_{0C}$ decreases while $\sigma_{0C}$ increases with $\phi_{CM}$ , following a power-law relationship. The relevant exponents are utilized for fractal scaling analysis, discussed next in this section. We note that the mature acid gels have higher moduli than the mature enzyme gels at all values of $\omega$ and $\gamma_0$ (comparing Figures 7A, 8A, and 9A with 7B, 8B, and 9B, respectively). Meanwhile, the mature acid gels yield and break at lower $\gamma_{0C}$ than that of mature enzyme gels (Figure 9C). This can be attributed to a more finely stranded network of acid versus enzyme gels. That can be due to the difference in the corresponding interactions dictated by the type of coagulant used for gelation. Overall, the $\phi_{CM}$-dependence is a power-law, which suggests a fractal network structure. We now discuss these features from the structural point of view.

**Fractal scaling with concentration of mature gels:** To extract a structural determination of gels from their rheology well beyond their gelation threshold requires models or so-called scaling theories. Such a framework focuses solely on the fully formed network.

**Shih model.** Based on fractal scaling concepts already available for polymer gels, Shih et al[3] proposed a scaling for colloidal gels that established a relationship between the mean size of the fractal aggregates/flocs $\xi_{flocs}$, and the overall volume fraction of particles $\phi$ (presumed equivalent to the particle volume fraction within the aggregate). Then, the fractal dimension $d_{f-flocs}$ is determined by,

$$\xi_{flocs} = a_{eff}.\phi^{\frac{1}{d_{f-flocs}-D}} \tag{10}$$

where $a_{eff}$ is the effective particle size, and $D$ $(= 3)$ is the Euclidean dimension of the system. The physical validity of this scaling (Eqn. 10) assumes $\phi \ll 1$ (i.e. below close packing), and presence of mass-fractal aggregates with voids and heterogeneity. The gel is essentially a collection of fractal flocs, each of which is also a weak linear spring. The scaling effectively describes two different types of gels, i.e., strong-link regime and weak-link regime. In strong-link regime, flocs are large, usually formed at low particle concentrations, with elastic behaviour of the gels dominated by elastic constant of the flocs ($K_\xi$). In weak-link regime, flocs are small, usually formed at higher particle concentrations, with elastic behaviour of the gels dominated by elastic constant of the inter-flock links ($K_l$). The weak links here are not chemical bonds but effective mechanical connections, and the corresponding network is more homogenous than the strong link case. The relevant expressions, as well as the deviation observed from this scaling are explained in the Section 5.1 of the SI.

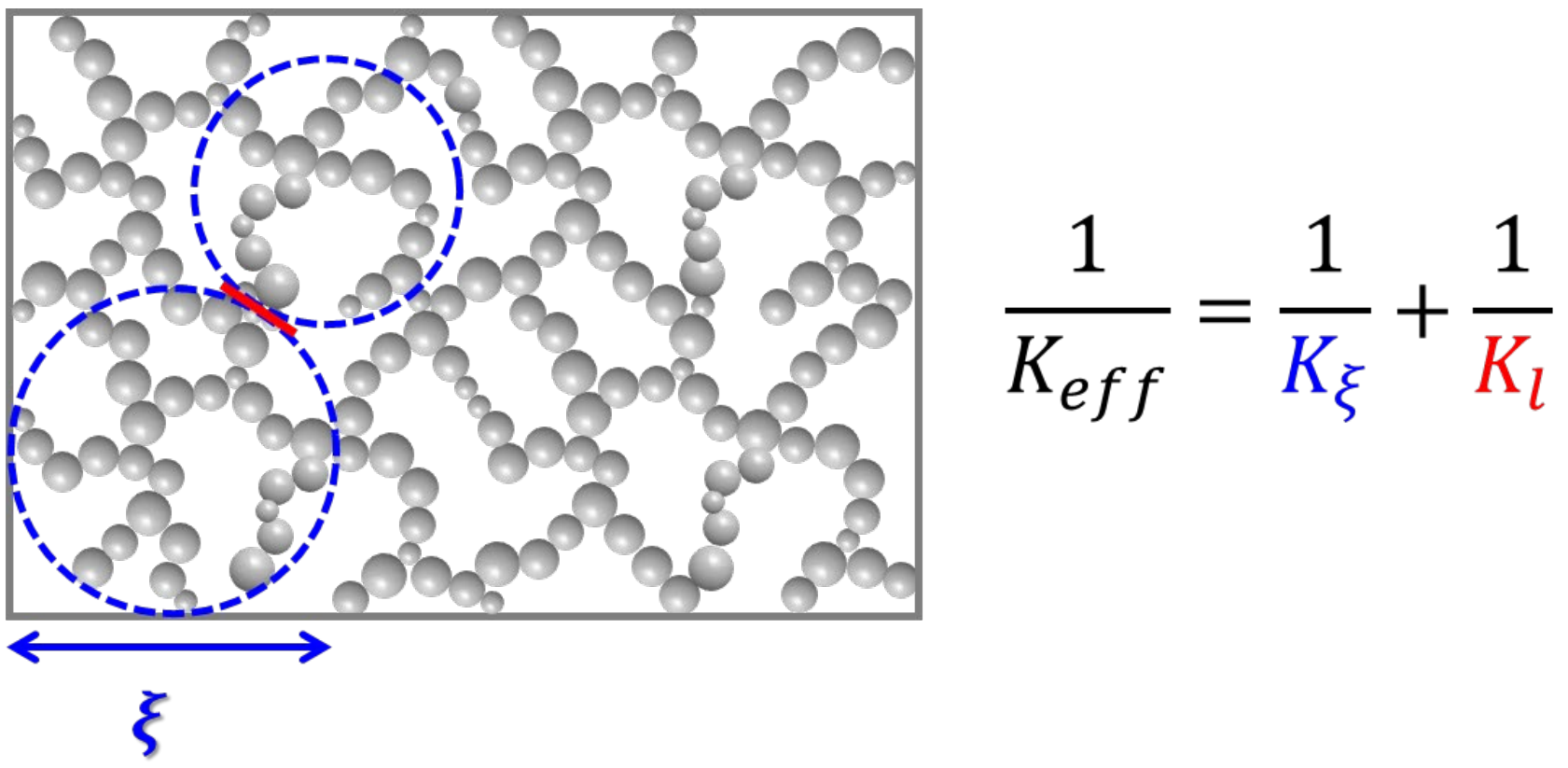

**Figure 10.** Schematic representation of the fractal structure of a colloidal gel based on the scaling theory of Wu & Morbidelli[68]. The red marks the "link" between two blue aggregates.

**Wu & Morbidelli model.** Wu & Morbidelli[68] introduce a transition between the strong-link and weak-link regimes of the Shih scaling[3], through an effective microscopic elastic constant ($K_{eff}$) combining $K_l$ and $K_\xi$ as depicted in Fig 9. This introduces a new, more empirical than physical, term $a$ which is 1 (weak link) or 0 (strong link). When 0 < a < 1, the gel is in the transition regime. The rheological properties of such a gel are related to $\varphi$ by,

$G' \sim \phi^{\frac{\beta}{D-\mathrm{df-flocs}}}$, $\gamma_{0C} \sim \phi^{\frac{D-\beta-1}{D-\mathrm{df-flocs}}}$, with

$$\beta = (D-2) + (2+x)(1-a) \qquad (11)$$

where $D$ $(=3)$ is the Euclidean dimension, and $x \in [1, 1.3]$ represents the backbone dimension or tortuosity of the network for colloidal gels.[68] We note that the assumptions for $\gamma_{0C}$ have stronger consequences than that for $G'$ in these scaling theories. While $G'$ scaling depends only on structural characteristics such as cluster fractality, the $\gamma_{0C}$ scaling relies on assuming a specific yielding mechanism, viz. fracture at the inter-floc links (weak-link gels) or within the flocs (strong-link gels), and a combination of both for the transition regime gels of Wu & Morbidelli.[68] For the mature gels of our work, we apply Wu & Morbidelli[68] scaling to the power-law exponents obtained from their $\phi_{CM}$-dependence of moduli ($G'_{FS}$, $G'_{AS}$), $G_0$, and critical strain $\gamma_{0C}$ (Figure 9), and present the corresponding parameters in Table 4.The scaling of the three quantities agree together to produce the same exponents.

**Table 4.** Wu & Morbidelli[68] scaling parameters for rheological properties of mature gels of present work. The fractal dimension $d_f$ is obtained (using Eqn. 11) from power-law fitting exponents of $G'$ and $\gamma_{0C}$ variation with $\phi$ (here $\phi_{CM}$) wherein $G'$ is either $G_0$, $G'_{FS}$, or $G'_{AS}$ (refer Figure 9). Here, $G'_{FS}$ is the linear moduli (at $\omega$ = 6.28 rad/s) from frequency sweep, $G_0$ is from modified weak gel model (Eqn. 9) fits of the frequency sweep data, and $G'_{AS}$ is the linear moduli (at $\gamma_0 = 1\%$) from amplitude sweep.

| Gel | Moduli | $d_{\mathrm{f-flocs}}$ | $x$ | a |
|---|---|---|---|---|
| Enzyme | $G_0$ | 2.517 ± 0.049 | 1, 1.3 | 0.59 ± 0.04, 0.63 ± 0.03 |
| | $G'_{FS}$ | 2.518 ± 0.052 | 1, 1.3 | 0.59 ± 0.04, 0.63 ± 0.03 |
| | $G'_{AS}$ | 2.499 ± 0.049 | 1, 1.3 | 0.59 ± 0.04, 0.63 ± 0.03 |
| Acid | $G_0$ | 2.431 ± 0.109 | 1, 1.3 | 0.48 ± 0.09, 0.53 ± 0.08 |
| | $G'_{FS}$ | 2.398 ± 0.132 | 1, 1.3 | 0.47 ± 0.11, 0.52 ± 0.09 |
| | $G'_{AS}$ | 2.465 ± 0.118 | 1, 1.3 | 0.49 ± 0.09, 0.54 ± 0.08 |

It is observed that the $d_{f-flocs}$ values remain the same irrespective of the $x$ value chosen for analysis, but the $a$ values differ. The mature enzyme gels and acid gels show $d_{f-flocs}$ ~ 2.51 and $d_{f-flocs}$ ~ 2.43 respectively. These values are close to the $d_f$ values obtained from the equilibrium USANS (Table 3). We also note that the mature enzyme gels are weaker ($a$ ~0.6) than that of the mature acid gels ($a$ ~0.5) as per the Wu & Morbidelli[68] scaling, in agreement with our prior rheological observations.

**Micro-elastic model:** Finally, the SI reports additional work on the gel elasticity model of Mellema et al recently applied to rennet-induced casein gels[31], similar to the model of Bremer at al[30], to describe the overall. The Shih et al[3] and Wu & Morbidelli[68] scaling theories are essentially coarse-grained versions of Mellema et al[31] model. Therefore, we use the $d_f$ values of our matured gels obtained from Wu & Morbidelli[68] scaling (Table 4) to explore the model of Mellema et al[31] (refer to section 5.2 of SI). Thus our mature gels can be defined as category IV-type particulate gels, similar to the casein gels of Mellema et al,[31] where fractal clusters are composed of straight, flexible strands that undergo stretching upon deformation during rheology.

## 4 Rationalization of Perspectives from USANS and Rheology

In this last section we will rationalize information extracted from USANS curves via comparison of a modelled scattering curve with data with the perspective taken from the analysis of oscillatory rheology of mature gels. The essential parameters, fractal scaling ($d_f$), and a correlation length-scale ($\xi$) for this scaling, are the same for each model, the rheological model is based on the mechanical properties of constituents and their connectivity, and the model for the USANS is a direct representation of the Fourier transform of the density correlation function.

**4.1 Fractal dimensions.** The fractal dimensions ($d_{f-USANS}$ *and* $d_{f-flocs}$) are summarized in Table 5. It is observed that the $d_{f-flocs}$ for mature acid gels are slightly less than that for enzyme gels. They are not very different, within the error bars, between USANS and rheology for each coagulation type.

**Table 5**. Fractal dimensions obtained from USANS and rheology of the matured gels of present work.

| Gel | Fractal dimension, $d_f$ | Technique |
|---|---|---|
| Enzyme | 2.503 ± 0.011 | USANS |
| | 2.511 ± 0.042 | rheology |
| Acid | 2.480 ± 0.009 | USANS |
| | 2.431 ± 0.102 | rheology |

**5.2 Length-scales.** There is a size range over which the fractal descriptions apply: $\xi_{USANS}$ is a consequence of the fitting of gel USANS data (Figure 4) with constraints defined by the fitting of the reconstituted skimmed milk (Figure 2 and Table 2); and $\xi_{Flocs}$ is calculated from Eqn. 10 with using values for $a_{eff}$ (equal to R, Figure 2) and $\phi_{CM}$( 0.101, Table 3). All the values, together with their relative errors, are plotted on Figure 10 with the USANS data with the y-axis $I_{slit}(q).q^{1.3}$ to accentuate deviation from linear fractal low-q behavior on the log (intensity) versus log q from the fractal network. The exponent 1.3 reflects RLCA behavior for slit smeared instrument. The region parallel to the X-axis is indicative of this behavior, with the deviation from this behavior at high q, associated with length-scales/density fluctuations at and below the radius of the CM (R), and in low-q, region the density fluctuations associated with the correlation length of the gel network. The ability to examine these length-scales is a strength of the neutron based Bonse-Hart double crystal diffractometer over contemporary pin-hole X-ray and neutron small angle scattering.[69] Strikingly, the correlation lengths calculated from the rheological outputs, $\xi_{floc}$, are considerably larger than those the determined from fitting of USANS data and outside the range of contemporary USANS data. In other terms, in the conceptual framework of the Wu & Morbidelli model[68] shown in Figure 10, that suggests that connective units for the elastic properties of the network are not the same as the density fluctuations which are related to USANS curves.

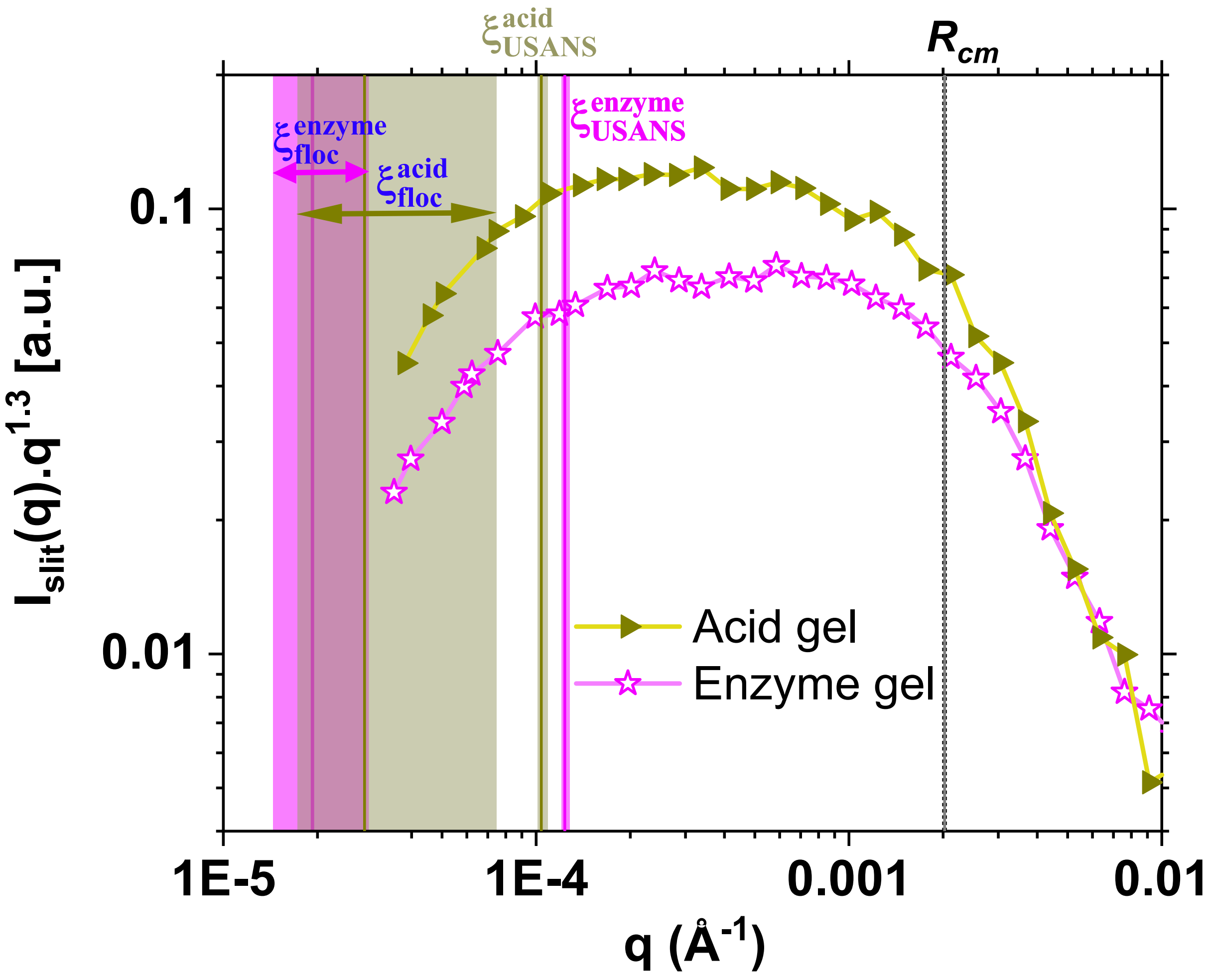

**Figure 11.** The scaled slit-smeared scattering intensity, $I_{slit}(q)$, from equilibrium USANS of the gels multiplied by $q^{1.3}$, as a function of the scattering vector, obtained at the largest time in enzymatic coagulation (pink stars) and acid coagulation (dark yellow triangles) of reconstituted skim milk at the representative concentration of 10 % w/w. Vertical lines and relative errors (shaded bars) to show: length scales associated with particle radius (1/R); correlation length from USANS modelling obtained from Chen and Teixeira[51, 52] fitting ($1/\xi_{USANS}$); and correlation length from rheology ($1/\xi_{Flocs}$). The two correlation lengths for acid and enzyme are labelled with a superscript.

We note from Eqn. 10 that the $\xi_{Rheo}$ of both our mature gels exhibit a power-law decrease with $\phi_{CM}$ (since $d_f < D(= 3)$). Hence, at higher $\phi_{CM}$, crowding limits the growth of large fractal clusters in the system which leads to compact structure. Consequently, the linear moduli measured during rheology of the matured gels also increase with $\phi_{CM}$ , as remarked earlier in section 3.2.2). In summary, this suggests that the fractal network structure of mature dairy gels may have a more complex internal hierarchy than a simple colloidal gels:

- a fractal arrangement of casein micelles of size R, revealed by USANS with an upper size limit, $\xi_{USANS}$ around 1 micron following the fractal scaling of RLCA-like aggregation of the CMs.
- a fractal arrangement of larger objects with an upper size limit of size $\xi_{Flocs}$, not observed directly but inferred from rheology in the mature state, with rather similar value of $d_{f\text{-}Flocs}$.

Importantly the sizes $\xi_{Flocs}$ and $\xi_{USANS}$ are revealed to be inequivalent. The former, obtained using rheology scaling analysis, are obtained the properties from the linear viscoelastic region summed over all contributions at the different scales.

A recent work[70] studies a material made of elementary objects sufficiently small, that their complete hierarchical description can be observed up to the largest sizes if using sufficiently small q angle scattering. These authors[70] studied gel formation after the flow cessation rheology of attractive carbon black suspensions. The extended SAXS reveals distinct features in the low-q, mid-q and high-q regions corresponding to the 3 length-scales $\xi_2$, $\xi_1$, and $a$, respectively. The hierarchical description involves 3-length-scales fractal scaling, clusters-of-cluster, and cluster interpenetration models, applicable when flow intervenes with colloidal gelation. In the clusters of cluster model, colloids of size $a$ tend to condense into clusters of size $\xi_1$ and fractal dimension $d_1$ which themselves form superaggregates of size $\xi_2$ and fractal dimension $d_2$. In the cluster interpenetration model, the colloids rather condense into large clusters of size $\xi_2$ and fractal dimension $d_2$ separated by a structural distance (i.e., centre-to-centre distance between two clusters) $\xi_1$.

We note that these models reduce into the common rheological scaling theories explored by our work, while, even by using the lowest accessible $q$ through USANS, we observe only a single correlation length in the system. This is due to the larger elementary particles in skim milk - CM, than in carbon black suspensions described above.

## 6. Summary and Conclusions

We have provided a systematic and comparative approach for fractal analysis of the networked gels using oscillatory rheology and ultra-small angle neutron scattering (USANS) studies on model dairy gels formed via enzymatic and acid coagulation of reconstituted skim milk. The length-scales explored by USANS are one decade larger than what formerly obtained through SAXS.[58] This range is shows an important overlap with range very relevant for rheology.

While our study permits to detect, by rheology, the build-up of elastic modulus via the cross point (G' > G") and a "critical gel-point" G'/G"" identical for all frequencies, most of the studies concern longer times for which gels undergo further aging, becoming "mature":

- USANS provides, for length-scales, up to 3 μm, a robust route to visualization of network formation. Initially, reconstituted milk, a simple model with hard sphere structure factor combined with the form factor of polydisperse spheres (log-normal distribution) is shown to be sufficient to describe the USANS. Growing of the network arises around 1 h at large q, develops to low q values until stabilization recorded after a few hours. A fractal structure factor is then essential to capture the low-q power-law scattering.
- Rheology also follows fractal scaling: a first one at gel point (1 h), which then evolves in a second one, in mature gels, most studied here. Following Wu & Morbidelli[68] such fully coagulated gel networks are described as composed soft, interconnected fractal flocs. The relevant gel strength is dictated by both inter-floc and intra-floc elasticity arising from the shear-induced stretching of underlying straight/ \flexible network strands.

A first fractal dimension 2.3 +/- 0.1 is shown by USANS from 30 min to establish in agreement with RLCA-like fractal scaling, until gels are mature, while the scaling theory by Wu & Morbidelli[68] allows to indirectly extract a value 2.4 +/- 0.1, which is close to the initial one. However, USANS reveals a correlation length $\xi_{USANS}$ ~< 1 μm while rheology scaling provides, indirectly, an estimate of a correlation length $\xi_{Flocs}$. More precisely, depending on the options chosen for the elementary unit of the flocs, $\xi_{Flocs}$ spans from close to $\xi_{USANS}$ to higher values, not resolvable by USANS. In this case, this would suggest a double fractal structure. Hence we can rationalize that the macroscopic elasticity (from rheology) contains contributions from the different scales, i.e. combines initial micro-structural evolution (seen from USANS) with another macroscale (> 1 μm) concomitant processes. This is akin to the two-step aging reported formerly.[55]

We finally note that this work permits us to compare enzymatic and acid gelation. Enzyme coagulation has faster structural kinetics (from USANS) but slower aging kinetics (from rheology), while the opposite is observed for gelation via acid coagulation. This manifests in the formation of weak but ductile mature gels from enzyme coagulation with $\xi_{Flocs}$ more widely separated from $\xi_{USANS}$ correlation lengths, $\xi_{Flocs}$ than for the mature gels from acid coagulation, which form strong but brittle gels with closely related correlation lengths.

## 7. Supplementary Material

The supplementary material contains ancillary rheological investigations referred to in the text.

## ACKNOWLEDGEMENTS

This project has received funding from the European Union's Horizon 2020 research and innovation program under the Marie Skłodowska-Curie grant agreement No. 101034266.

USANS beamtime was granted by ANSTO based on a peer-reviewed proposal submission process. Rheology instrument access was provided by Dr. Henrich Frielinghaus at the Forschungszentrum Julich, MLZ Garching campus.

## REFERENCES

1. J. A. Lucey, International Journal of Dairy Technology **57** (2-3), 77-84 (2004).
2. J. A. Lucey, Journal of Dairy Science **85** (2), 281-294 (2002).
3. W. H. Shih, W. Y. Shih, S. I. Kim, J. Liu and I. A. Aksay, Physical Review A **42** (8), 4772-4779 (1990).
4. Z. Yang, L. Cheng, L. de Campo, E. P. Gilbert, R. Mittelbach, L. Luo, A. Ye, S. Li and Y. Hemar, Food Hydrocolloids **137**, 108381 (2023).
5. L. F. van Heijkamp, I. M. de Schepper, M. Strobl, R. H. Tromp, J. R. Heringa and W. G. Bouwman, Journal of Physical Chemistry A **114** (7), 2412-2426 (2010).
6. N. Callaghan-Patrachar, F. Peyronel, D. A. Pink, A. G. Marangoni and C. P. Adams, Food Hydrocolloids **116**, 106622 (2021).
7. C. P. Adams, N. Callaghan-Patrachar, F. Peyronel, J. Barker, D. A. Pink and A. G. Marangoni, Food Structure **21**, 100120 (2019).
8. A. Bouchoux, J. Ventureira, G. Gésan-Guiziou, F. Garnier-Lambrouin, P. Qu, C. Pasquier, S. Pézennec, R. Schweins and B. Cabane, Soft Matter **11** (2), 389-399 (2015).
9. C. G. De Kruif, Journal of Applied Crystallography **47** (5), 1479-1489 (2014).
10. C. Holt, C. G. de Kruif, R. Tuinier and P. A. Timmins, Colloid Surf. A-Physicochem. Eng. Asp. **213** (2-3), 275-284 (2003).
11. B. Ingham, A. Smialowska, G. D. Erlangga, L. Matia-Merino, N. M. Kirby, C. Wang, R. G. Haverkamp and A. J. Carr, Soft Matter **12** (33), 6937-6953 (2016).
12. D. Lehner, G. Kellner, H. Schnablegger and O. Glatter, Journal of Colloid and Interface Science **201** (1), 34-47 (1998).
13. D. Lehner, P. Worning, G. Fritz, L. Øgendal, R. Bauer and O. Glatter, Journal of Colloid and Interface Science **213** (2), 445-456 (1999).
14. J. P. Mata, P. Udabage and E. P. Gilbert, Soft Matter **7** (8), 3837-3843 (2011).
15. A. Rohart, C. Michon, J. Confiac and V. Bosc, Dairy Science & Technology **96** (4), 459-475 (2016).
16. P. H. Stothart and D. J. Cebula, Journal of Molecular Biology **160** (2), 391-395 (1982).
17. R. H. Tromp and W. G. Bouwman, Food Hydrocolloids **21** (2), 154-158 (2007).
18. E. D. Gado, D. Fiocco, G. Foffi, S. Manley, V. Trappe and A. Zaccone, in *Fluids, Colloids and Soft Materials* (2016), pp. 279-291.
19. P. Meakin, Adv. Colloid Interface Sci. **28**, 249-331 (1987).
20. S. Lazzari, L. Nicoud, B. Jaquet, M. Lattuada and M. Morbidelli, Adv. Colloid Interface Sci. **235**, 1-13 (2016).
21. J. S. Pedersen, T. L. Møller, N. Raak and M. Corredig, Soft Matter **18** (45), 8613-8625 (2022).
22. C. G. de Kruif, T. Huppertz, V. S. Urban and A. V. Petukhov, Adv. Colloid Interface Sci. **171-172**, 36-52 (2012).
23. D. J. McMahon and R. J. Brown, Journal of Dairy Science **67** (3), 499-512 (1984).

24. C. G. De Kruif and C. Holt, in *Advanced Dairy Chemistry—1 Proteins: Part A / Part B*, edited by P. F. Fox and P. L. H. McSweeney (Springer US, Boston, MA, 2003), pp. 233-276.
25. C. G. de Kruif and E. B. Zhulina, Colloids and Surfaces A: Physicochemical and Engineering Aspects **117** (1), 151-159 (1996).
26. M. Corredig, P. K. Nair, Y. Li, H. Eshpari and Z. Zhao, Journal of Dairy Science **102** (6), 4772-4782 (2019).
27. U. Bonse and M. Hart, Applied Physics Letters **7** (9), 238-& (1965).
28. M. Bantawa, B. Keshavarz, M. Geri, M. Bouzid, T. Divoux, G. H. McKinley and E. Del Gado, Nature Physics (2023).
29. M. Muthukumar, Macromolecules **22** (12), 4656-4658 (1989).
30. L. G. B. Bremer, T. van Vliet and P. Walstra, Journal of the Chemical Society, Faraday Transactions 1: Physical Chemistry in Condensed Phases **85** (10), 3359-3372 (1989).
31. M. Mellema, J. H. J. van Opheusden and T. van Vliet, Journal of Rheology **46** (1), 11-29 (2002).
32. C. Rehm, L. de Campo, A. Brule, F. Darmann, F. Bartsch and A. Berry, Journal of Applied Crystallography **51** (1) (2018).
33. O. Glatter and O. Kratky, *Small angle x-ray scattering*. (Academic Press, 1982).
34. L. de Campo, C. J. Garvey, C. D. Muzny, C. Rehm and H. J. M. Hanley, Journal of Colloid and Interface Science **533**, 136-143 (2019).
35. N. Xiong, D. Mannicke, T. Lam and N. Hauser, (ANSTO, 2017).
36. S. R. Kline, Journal of Applied Crystallography **39**, 895-900 (2006).
37. A. Otsuki, L. de Campo, C. J. Garvey and C. Rehm, Colloids and Interfaces **2** (3) (2018).
38. J. S. Pedersen, Journal De Physique Iv **3** (C8), 491-498 (1993).
39. M. Doucet, J. Cho, G. Alina, J. Bakker, W. Bouwman, P. Butler, K. Campbell, M. Gonzales, R. Heenan, A. Jackson, P. Juhas, S. King, P. Kienzle, J. Krzywon, A. Markvardsen, T. Nielsen, L. O'Driscoll, W. Potrzebowski, F. L. R., T. Richter, P. Rozycko and A. Washington, (2017).
40. M. Kotlarchyk and S. H. Chen, The Journal of Chemical Physics **79** (5), 2461-2469 (1983).
41. J. S. Higgins and H. Benoit, *Polymers and neutron scattering*. (Clarendon Press, Oxford, 1994).
42. G. N. Smith, Food Hydrocolloids **113**, 106511 (2021).
43. G. N. Smith, E. Brok, M. V. Christiansen and L. Ahrné, Soft Matter **16** (43), 9955-9963 (2020).
44. A. E. Whitten, S. Cai and J. Trewhella, Journal of Applied Crystallography **41** (1), 222-226 (2008).
45. A. Lopez-Rubio and E. P. Gilbert, Trends in Food Science & Technology **20** (11), 576-586 (2009).
46. R. Jenness, E. O. Herreid, W. J. Caulfield, L. H. Burgwald, E. L. Jack and S. L. Tuckey, Journal of Dairy Science **25** (11), 949-960 (1942).
47. S. Nöbel, C. Kern, A. Sonne, B. Bähler and J. Hinrichs, International Dairy Journal **59**, 80-84 (2016).
48. M. Mours and H. H. Winter, Rheologica Acta **33** (5), 385-397 (1994).
49. J. K. Percus and G. J. Yevick, Physical Review **110** (1), 1-13 (1958).
50. S. V. G. Menon, C. Manohar and K. S. Rao, The Journal of Chemical Physics **95** (12), 9186-9190 (1991).
51. J. Teixeira, Journal of applied crystallography **21**, 781-785 (1988).
52. S.-H. Chen and J. Teixeira, Physical Review Letters **57** (20), 2583-2586 (1986).

53. H. H. Winter, Polymer Engineering & Science **27** (22), 1698-1702 (1987).
54. H. H. Winter and F. Chambon, Journal of Rheology **30** (2), 367-382 (1986).
55. M. A. V. Axelos and M. Kolb, Physical Review Letters **64** (12), 1457-1460 (1990).
56. D. Durand, M. Delsanti, M. Adam and J. M. Luck, Europhys. Lett. **3** (3), 297 (1987).
57. C. Michon, G. Cuvelier and B. Launay, Rheol. Acta **32** (1), 94-103 (1993).
58. J. Bauland, G. Manna, T. Divoux and T. Gibaud, Physical Review Materials **8** (7), L072601 (2024).
59. S. Ben-Harb, M. Panouillé, D. Huc-Mathis, G. Moulin, A. Saint-Eve, F. Irlinger, P. Bonnarme, C. Michon and I. Souchon, Food Hydrocolloids **77**, 75-84 (2018).
60. W. Xiuju and Z. Zhengtao, in *Current Issues and Advances in the Dairy Industry*, edited by A. I. Salam (IntechOpen, Rijeka, 2022), pp. Ch. 5.
61. M. Geri, B. Keshavarz, T. Divoux, C. Clasen, D. J. Curtis and G. H. McKinley, Physical Review X **8** (4), 041042 (2018).
62. R. C. Koeller, Journal of Applied Mechanics **51** (2), 299-307 (1984).
63. D. Gabriele, B. de Cindio and P. D'Antona, Rheol. Acta **40** (2), 120-127 (2001).
64. L. Z. Rogovina and G. L. Slonimskii, Russian Chemical Reviews **43** (6), 503 (1974).
65. L. Bohlin, Journal of Colloid and Interface Science **74** (2), 423-434 (1980).
66. K. A. Ramya, R. Srinivasan and A. P. Deshpande, Rheol. Acta **57** (2), 181-195 (2018).
67. K. Hyun, M. Wilhelm, C. O. Klein, K. S. Cho, J. G. Nam, K. H. Ahn, S. J. Lee, R. H. Ewoldt and G. H. McKinley, Progress in Polymer Science **36** (12), 1697-1753 (2011).
68. H. Wu and M. Morbidelli, Langmuir **17** (4), 1030-1036 (2001).
69. A. Magerl, H. Lemmel, M. Appel, M. Weisser, U. Kretzer and M. Zobel, Journal of Applied Crystallography **57** (5), 1282-1287 (2024).
70. L.-V. Bouthier and T. Gibaud, Journal of Rheology **67** (3), 621-633 (2023).